\documentclass[final,5p,authoryear,times]{elsarticle} 
\pdfoutput=1

\usepackage{graphicx}
\usepackage{amsmath}
\usepackage{booktabs}
\usepackage{colortbl} 
\usepackage{amssymb}
\usepackage{makecell}
\usepackage{float}
\usepackage{color}
\newcommand{\rthis}[1]{\textcolor{black}{#1}}
\usepackage[T1]{fontenc}
\usepackage{ae,aecompl}
\usepackage{enumitem}
\usepackage{times}
\usepackage[hidelinks]{hyperref}
\usepackage{xcolor}
\usepackage{subfig}
\usepackage[ruled,vlined]{algorithm2e}
\usepackage[utf8]{inputenc}
\usepackage{tabularx}
\usepackage{amsmath}
\usepackage{adjustbox}
\usepackage{graphicx}
\usepackage{float}
\usepackage{placeins}

\hypersetup{
    colorlinks,
    linkcolor={red!50!black},
    citecolor={blue!50!black},
    urlcolor={blue!80!black}
}
\usepackage{multirow}
\usepackage{mwe}
\usepackage{rotating}
\usepackage[normalem]{ulem}
\useunder{\uline}{\ul}{}
\usepackage{dblfloatfix}
\usepackage{microtype}

\DeclareUnicodeCharacter{2212}{-}
\journal{Astronomy \& Computing}

\begin{document} \sloppy
\begin{frontmatter}

\title{Comparison of symbolic regression algorithms in Star/galaxy/quasar separation}

\author[1]{Rachit Deshpande}\ead{ai24mtech11003@iith.ac.in}
\author[2]{Shantanu  Desai}\ead{shantanud@phy.iith.ac.in}
\address[1]{Department of  Artificial Intelligence, IIT Hyderabad, Kandi,  Telangana-502284, India}
\address[2]{Department of Physics, IIT Hyderabad, Kandi,  Telangana-502284, India}

\begin{abstract} 
\label{abstract}
This work investigates symbolic regression (SR) as an interpretable alternative to black-box machine learning for the classification of stars, galaxies, and quasars in the Sloan Digital Sky Survey Data Release 17 (SDSS DR17). We conduct a systematic comparative study of four state-of-the-art SR frameworks: {\tt PySR}, Exhaustive Symbolic Regression ({\tt ESR}) with MDL-based selection, Physical Symbolic Optimization ({\tt PhySO}) using deep reinforcement learning, and Multi-View Symbolic Regression ({\tt MvSR}). By deriving compact analytic functions (complexity $\leq$ 10) on a representative training subset and subsequently evaluating them via \rthis{an 80,000-sample 5-fold cross-validation threshold optimization phase and a subsequent 10,000-sample unseen hold-out test set}, we map spectroscopic redshift ($z$) to continuous classification scores. Our results demonstrate that these low-complexity expressions achieve high predictive reliability, with {\tt MvSR} reaching a \rthis{cross-validation Cohen’s Kappa of 0.8956 (0.8876 on the hold-out set)} and {\tt PhySO} achieving exceptional parametric stability ($\sigma < 0.002$). \rthis {We note however that the resulting equations returned by Symbolic regression are purely empirical and no physical significance should be ascribed to these equations.}
\end{abstract}

\begin{keyword}
Symbolic Regression, Astroinformatics, Spectroscopic Classification, Star-Galaxy-Quasar Separation, Deep Reinforcement Learning, Explainable AI (XAI), Genetic Programming, Information Theory, Interpretable Machine Learning, Scientific Machine Learning (SciML), SDSS DR17.
\end{keyword}
\end{frontmatter}

\section{INTRODUCTION} 
\label{INTRODUCTION}
\par 
Photometric classification of stars, galaxies, and quasars from large-scale astronomical surveys  underpins a wide range of analyses in cosmology, galaxy evolution, and active galactic nuclei studies~\citep{Souma}. A large number of machine learning approaches, particularly deep neural networks and gradient-boosted trees have been applied to the problem of object classification using data from photometric surveys such as Blanco Cosmology Survey, Sloan Digital Sky Survey (SDSS), Dark Energy Survey etc~\citep{Desai12,Nacho,Chaini,Srinadh,Slater,Logan} (and references therein). Some of these achieve classification accuracies exceeding 95\%~\citep{Solorio23} on SDSS DR17~\citep{DR17}. However, these ``black box'' models obscure the underlying astrophysical relationships-such as redshift evolution and population density gradients-that govern object separability~\citep{hassija2024}.

To ameliorate this problem, symbolic regression (SR) provides a compelling alternative by jointly optimizing predictive performance and mathematical interpretability~\citep{Bartlett,Desmond25}. Rather than learning opaque weight matrices, SR searches directly for analytic expressions that map input features to continuous class scores. These scores can subsequently be thresholded into discrete labels, providing a transparent decision manifold. Recent advances have produced a diverse set of SR methodologies: stochastic genetic programming ({\tt PySR})~\citep{pysr}, deterministic exhaustive enumeration ({\tt ESR})~\citep{ESR}, reinforcement learning-guided methods with unit constraints ({\tt PhySO})~\citep{Tenachi}, and ensemble-style strategies like Multi-View Symbolic Regression ({\tt MvSR})~\citep{Russeil}. These have recently been applied to a variety of problems in Astrophysics and Cosmology (See ~\citealt{Patra,Darc,Martin26,Sousa-Neto} for a non-exhaustive list of recent applications of symbolic regression). \rthis{We numerate some of these recent applications. It was used to rediscover planetary laws of motion from trajectories of solar system objects~\citep{Lemos}. It was also used to determine   SZ scaling relations in galaxy clusters after incorporating feedback~\citep{Wadekar}. It has been used to improve models   of halo occupation distribution~\citep{Delgado}.  Some other applications include determination of 
neutron star properties~\citep{Patra}, reconstruction of dark matter profiles using weak lensing~\citep{Martin}, estimation of stellar masses from photometry and redshifts~\citep{Adarsh}, estimating neutron star radii from gravitational wave measurements~\citep{Bejger}, relation between black hole mass and galaxy properties~\citep{Jin}, constructing inflationary potentials etc~\citep{Sousa}. }

About a year ago, ~\citet{Fabio25} applied the {\tt  PySR} symbolic regression algorithm to SDSS DR17 data to derive an analytical expression which was used to separate stars, galaxies, and quasars. A genetic algorithm was also used to optimize the hyperparameters. The resulting combination achieved a Cohen Kappa value of  0.81.

We extend the analysis in ~\citet{Fabio25} and consider four complementary SR approaches to the same problem of star/galaxy/quasar classification on the same SDSS DR17 data, under a unified \rthis{80,000-sample} 5-fold stratified cross-validation protocol \rthis{and a subsequent 10,000-sample unseen hold-out test phase. Since ~\citet{Fabio25} had found that the spectroscopic redshift ($z$) is the best discriminator}, we first restrict  the input space by considering only $z$. \rthis{In the Appendix we consider the full photometric features.}
We then isolate and evaluate the capacity of each method to discover compact redshift-centric decision functions. Our analysis focuses not only on classification metrics such as Cohen’s Kappa and Balanced Accuracy-where we achieve \rthis{hold-out test results exceeding 0.88 and 91\%}, respectively-but also on the parametric stability of the optimized decision thresholds ($t_1$ and $t_2$).

This manuscript is structured as follows. The dataset is described in Sect.~\ref{sec:dataset}. The methodology is described in Sect.~\ref{sec:method}, results in Sect.~\ref{sec:results}, and we conclude in Sect.~\ref{sec:conclusions}.

\section{Dataset}
\label{sec:dataset}
We analyze the stellar classification catalog from SDSS DR17, which contains 100,000 spectroscopically confirmed objects labeled as galaxies, stars, or quasars. After removing the entries with missing redshift or class values and standardizing labels (mapping ``QUASAR'' to ``QSO''), the final working sample consists of 59,445 galaxies (59.45\%), 21,594 stars (21.59\%), and 18,961 quasars (18.96\%).

\begin{figure}[!t]
    \centering
    \includegraphics[width=0.9\columnwidth]{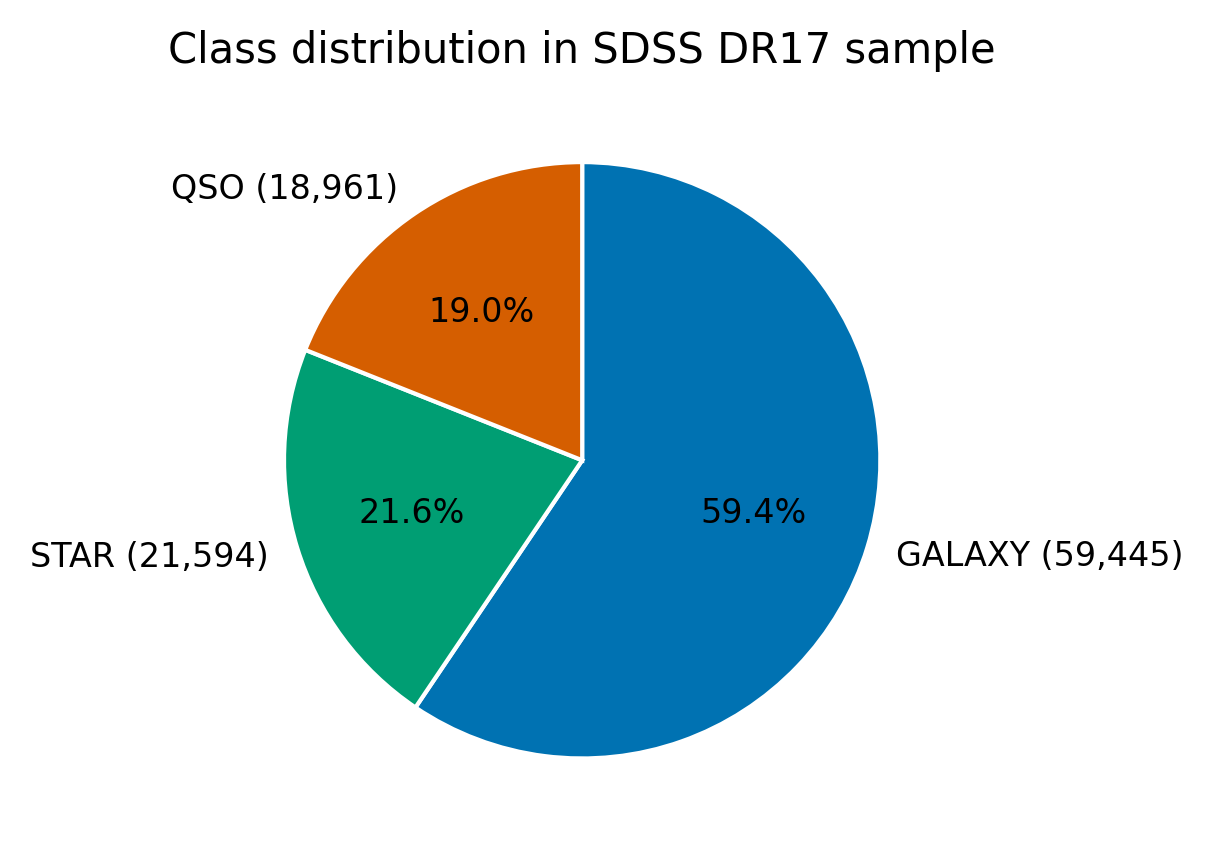}
    \caption{Class distribution of the SDSS DR17 dataset across GALAXY, STAR, and QSO categories.}
    \label{F1}
\end{figure}

Figure~\ref{F1} illustrates the pronounced class imbalance in the dataset, with galaxies forming the clear majority. Such imbalance presents well-known challenges for machine learning classifiers~\citep{Bethapudi}, particularly when evaluating performance on the underrepresented QSO population.
For the symbolic discovery phase, a representative 10\% subset (10,000 objects) of the dataset was utilized (similar to ~\citealt{Fabio25}) to ensure computational efficiency while preserving population diversity and class proportions. \rthis{The resulting equations were subsequently evaluated using a rigorous two-stage protocol: an 80,000-sample subset was utilized for 5-fold cross-validation to optimize decision thresholds, and a final 10,000-sample unseen hold-out set was reserved for global generalization testing.}

\begin{figure}[!t]
    \centering
    \includegraphics[width=0.9\columnwidth]{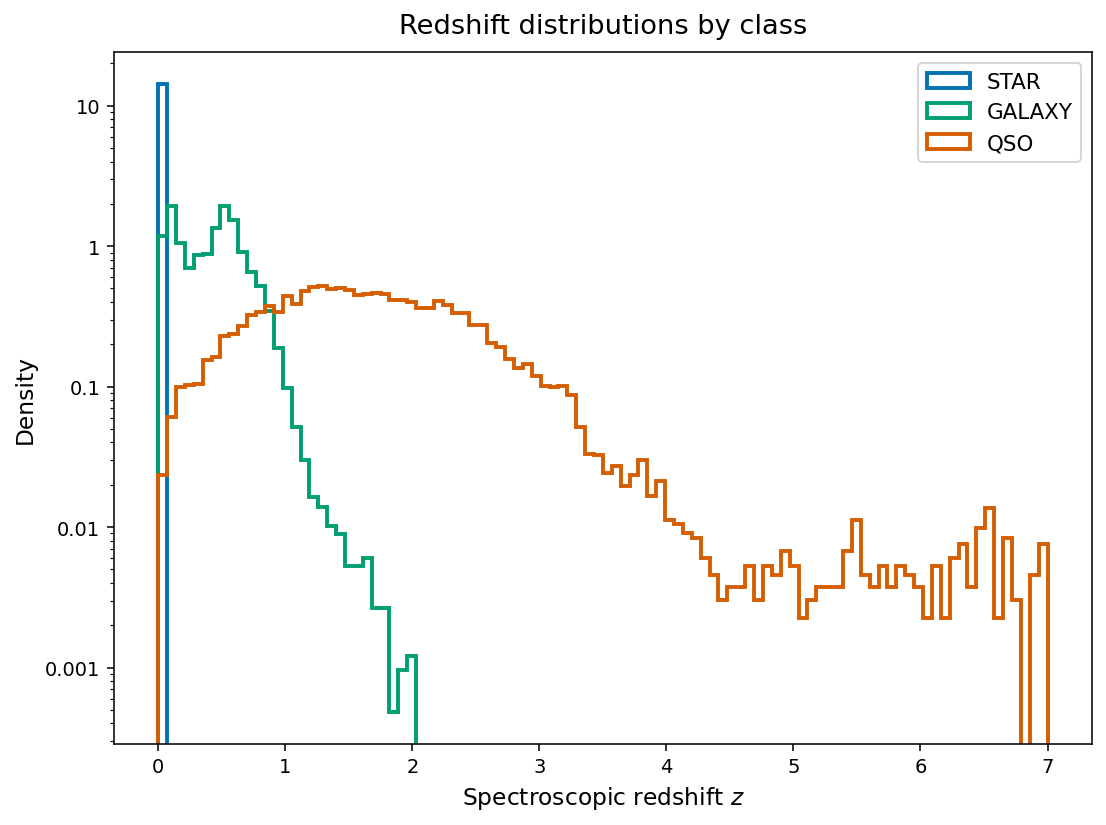}
    \caption{Normalized redshift distributions of stars, galaxies, and quasars shown on a logarithmic scale.}
    \label{F2}
\end{figure}

The redshift distributions shown in Figure~\ref{F2}, presented as normalized histograms with a logarithmic $y$-axis, exhibit characteristic population signatures that strongly motivate redshift-only classification. Stars concentrate in a narrow peak at $z \approx 0$, corresponding primarily to Galactic foreground objects. Galaxies form a broad distribution peaking between $z \sim 0.1$ and $z \sim 0.8$, typical of luminous red galaxies in this spectroscopic sample. Quasars display an extended high-redshift tail beyond $z \sim 2$, reaching values exceeding $z > 6$.

\begin{figure}[!t]
    \centering
    \includegraphics[width=0.9\columnwidth]{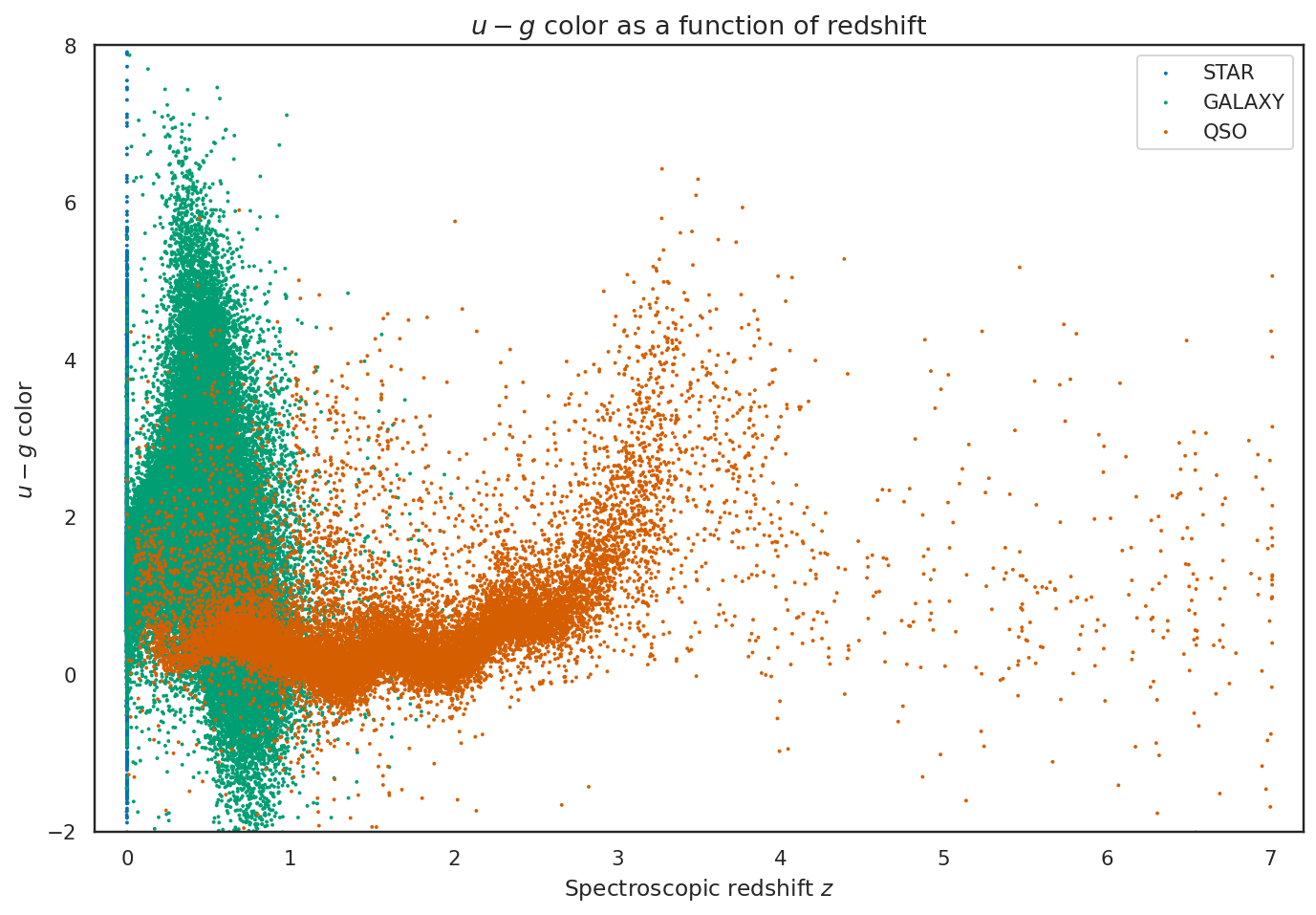}
    \caption{$u - g$ color as a function of redshift for stars, galaxies, and quasars, highlighting population-specific trends.}
    \label{F3}
\end{figure}

Figure~\ref{F3} shows the $u - g$ color--redshift relation, revealing systematic population differences. Stars follow the well-known stellar locus with minimal redshift evolution, while galaxies occupy a relatively stable intermediate-color region across the observed range~\citep{Desai12}. Quasars, however, exhibit pronounced color evolution; they typically appear blue (exhibiting a UV-excess with $u - g \lesssim 1$) at low redshifts, but transition to significantly redder colors ($u - g \gtrsim 2$) at $z > 2.5$ as Lyman-$\alpha$ forest absorption enters the SDSS $u$-band~\citep{Richards02}.

These empirical distributions provide strong motivation for developing compact symbolic regression models using redshift alone. The three populations occupy systematically distinct regions of this one-dimensional parameter space, suggesting that simple, mathematically concise analytic functions $f(z)$ can achieve meaningful class separation without the need for high-dimensional color spaces.

\rthis{It is important to note that because this study utilizes strictly spectroscopically confirmed objects, the measurement uncertainties associated with the input redshift feature are exceptionally small (typically $\sigma_z \approx 10^{-4}$ to $10^{-3}$). At this level of precision, the instrumental error is orders of magnitude smaller than the physical redshift spread of the targeted populations. Consequently, explicitly propagating observational uncertainties through the symbolic regression search algorithms was deemed unnecessary, as such micro-variations are insufficient to perturb the macroscopic algebraic decision boundaries discovered by the models.}

\section{Methodology}
\label{sec:method}

\subsection{General Approach}

Our methodology seeks to construct interpretable decision functions that map spectroscopic redshift to astrophysical class membership. Specifically, we learn continuous score functions $s(z)$ that take the  redshift $z$ as input and produce a scalar score predictive of the underlying object class. Two optimized thresholds, $t_1$ and $t_2$ (where $t_1 < t_2$), are used to convert the continuous score into discrete class labels. This process follows the decision manifold described in the following equation:
\begin{equation}
\label{eq:decision_manifold}
\mathrm{Class}(z) =
\begin{cases}
\text{GALAXY} & \text{if } s(z) < t_1 \\[4pt]
\text{STAR}   & \text{if } t_1 \le s(z) \le t_2 \\[4pt]
\text{QSO}    & \text{if } s(z) > t_2
\end{cases}
\end{equation}

This formulation allows each symbolic regression method to express class separation through a compact analytic function,  while preserving a transparent and physically interpretable decision structure. Unlike traditional black-box classifiers that rely on high-dimensional softmax layers, this one-dimensional scoring manifold forces the frameworks to discover the natural ordering of astrophysical populations along the redshift axis-beginning with nearby Galactic stars at $z \approx 0$ and extending to high-redshift quasars at $z > 2$.

By restricting the input to a single physically meaningful parameter, we isolate the representational capacity of each symbolic framework and evaluate its ability to capture the non-linear boundaries between populations using mathematically concise expressions.

\subsection{Experimental Protocol and Data Preprocessing}
\label{secpre}

The SDSS DR17 sample used in this study consists of 100,000 spectroscopically confirmed objects after standardizing class labels (mapping “QUASAR” to “QSO”)~\citep{DR17}. The final class distribution is GALAXY (59.45\%), STAR (21.59\%), and QSO (18.96\%). No entries with missing redshift or class labels were present in the cleaned dataset.

To maintain computational efficiency during the symbolic discovery phase-particularly for exhaustive symbolic regression ({\tt ESR}) and genetic search frameworks such as {\tt PySR} and {\tt MvSR}-a representative 10\% subset (10,000 objects) is initially sampled from the entire dataset, similar to that done in ~\citet{Fabio25}. This static subset enables deep exploration of the functional search space, while preserving the underlying astrophysical population density and class balance, allowing the models to discover a universal scoring function $s(z)$.

\rthis{Once the scoring functions are determined, we adopt a rigorous two-stage evaluation protocol to ensure statistical robustness and prevent data leakage. The remaining 90,000 objects are partitioned into an 80,000-sample optimization set and a 10,000-sample hold-out test set. We perform a 5-fold stratified cross-validation exclusively on the 80,000-sample optimization set to calibrate the decision thresholds. Finally, the generalized performance of these equations is evaluated on the 10,000-sample unseen hold-out set.}

A critical component of our unified protocol is the automated grid-search algorithm used to calibrate the decision thresholds $t_1$ and $t_2$. For every discovered scoring function $s(z)$, the algorithm searches candidate threshold pairs within the range defined by the 5th and 95th percentiles of the training fold scores. The pair $(t_1, t_2)$ that maximizes Cohen’s $\kappa$ coefficient (cf. Eq.~\ref{eq:cohenkappa}) is selected and subsequently evaluated on the unseen test set.

\begin{figure*}[htbp]
    \centering
    \includegraphics[width=1.2\columnwidth]{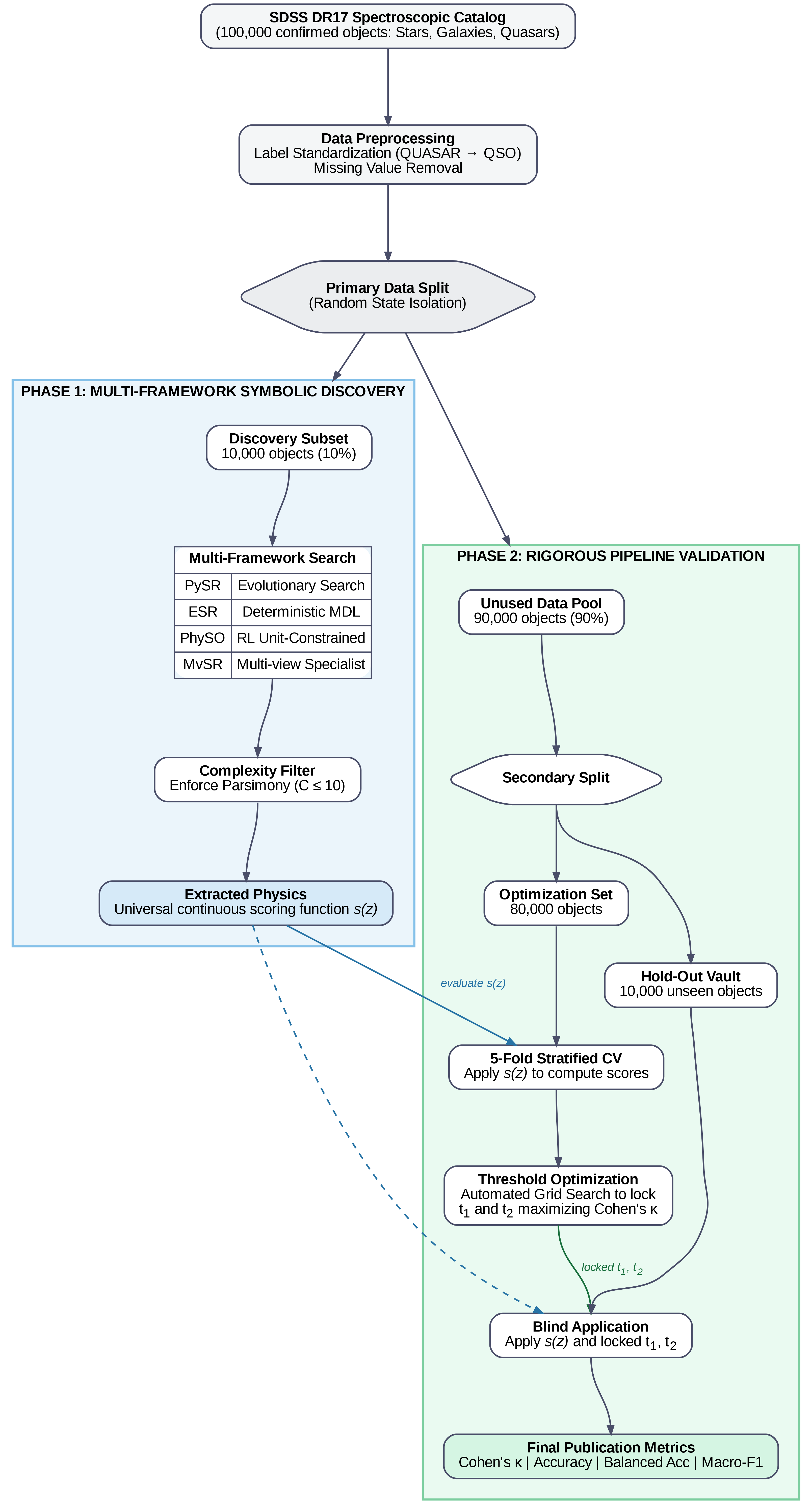}
    \caption{Unified experimental pipeline for SDSS DR17 classification.}
    \label{F4}
\end{figure*}

This procedure ensures that classification boundaries are optimally tuned for each symbolic framework under identical conditions, providing a fair benchmark of their respective mathematical search strategies.
The \rthis{optimization} performance metrics are reported as mean values with standard deviations \rthis{across the five CV folds, alongside the final deterministic performance on the unseen hold-out set}. The complete end-to-end pipeline, from symbolic discovery on the representative subset \rthis{to threshold calibration and final hold-out validation}, is summarized in Figure ~\ref{F4}.

\subsection{Performance Metrics and Evaluation Criteria}

The evaluation of the symbolic regression (SR) frameworks is conducted through a multi-tiered diagnostic process. Because these models are required to perform a classification task using a regressive search engine, we distinguish between Functional Fidelity (regression accuracy) and Classification Reliability (the success of the population separation).

\subsubsection{Functional Discovery and Structural Constraints}

During the initial symbolic search, the input spectroscopic redshift ($z$) is mapped to a target ordinal encoding. The following criteria govern this discovery phase:

\paragraph{Mean Squared Error (MSE):}
The primary loss function used during structural evolution is the MSE, which penalizes the squared distance between the continuous symbolic output $s(z_i)$ and the true target ordinal label $y_i$:
\begin{equation}
MSE = \frac{1}{N} \sum_{i=1}^{N} (y_i - s(z_i))^2 ,
\end{equation}
where $N$ represents the total number of evaluated samples. In this study, MSE ensures the topological preservation of the dataset. By using an ordinal scale, we force the symbolic engine to discover functions that naturally increase or decrease along the redshift manifold.

\paragraph{Structural Complexity ($C$):}

In accordance with Occam’s Razor, we enforce a complexity constraint to prevent structural over-fitting. Following standard practices in genetic programming \citep{pysr} and the methodological baseline established for this dataset \citep{Fabio25}, complexity is explicitly defined as the total node count ($|T|$) of the expression tree:

\begin{equation}
C(T) = \sum (\text{operators} + \text{variables} + \text{constants}) \leq 10
\end{equation}

Restricting $C \leq 10$ ensures that the resulting equations remain human-readable and mathematically parsimonious.

\subsubsection{Classification Reliability and Statistical Agreement}

After a scoring function $s(z)$ is finalized, it is discretized using two optimized thresholds, $t_1$ and $t_2$. To evaluate the success of this discretization, we employ metrics that are robust against class imbalance:

\paragraph{Cohen’s Kappa ($\kappa$):}

Unlike raw accuracy, Cohen’s $\kappa$ measures the agreement between the model and the true labels while correcting for chance~\citep{Cohen60} and is defined as follows:
\begin{equation}
\kappa = \frac{p_o - p_e}{1 - p_e} ,
\label{eq:cohenkappa}
\end{equation}
where $p_o$ is the observed accuracy and $p_e$ is the probability of random agreement, calculated as:

\begin{equation}
p_e = \frac{1}{N^2} \sum_{k} n_{k1} n_{k2}
\end{equation}

where $n_{k1}$ and $n_{k2}$ are the marginal totals for class $k$ in the confusion matrix.

\paragraph{Macro-Averaged F1-Score:}
The F1-score is the harmonic mean of precision and recall. We utilize the Macro-average to provide an unweighted mean of per-class performance:
\begin{equation}
F1_{k} = \frac{2 \cdot \text{Precision}_k \cdot \text{Recall}_k}{\text{Precision}_k + \text{Recall}_k}
\end{equation}
The macro average ($F1_{\text{macro}}$) is defined as follows:
\begin{equation}
F1_{\text{macro}} = \frac{1}{K} \sum_{k=1}^{K} F1_k,
\end{equation}
where $K=3$ is the total number of astrophysical classes.

\paragraph{Balanced Accuracy:}
Calculated as the arithmetic mean of class-specific recall values, this metric provides an unbiased accuracy score:
\begin{equation}
\text{Balanced Acc.} = \frac{1}{K} \sum_{k=1}^{K} \frac{TP_k}{TP_k + FN_k} ,
\end{equation}
where $TP$ and $FN$ are True Positives and False Negatives, respectively.

\subsubsection{Physical Failure Analysis}

To provide a deeper astrophysical interpretation, we utilize diagnostic tools that map mathematical errors to physical redshift regions:

\paragraph{Binned Recall Profiles ($R(z)$):}

By calculating recall within narrow redshift windows $\Delta z$, we identify ``Physical Transition Zones.''

\begin{equation}
R(z)_k = \frac{TP_k(z \in \Delta z)}{TP_k(z \in \Delta z) + FN_k(z \in \Delta z)}
\end{equation}

This identifies where the spectral signatures of populations become indistinguishable to a 1D model.

\paragraph{Kappa Sensitivity Heatmaps:}

We generate 2D grid visualizations of $\kappa$ across a range of $t_1$ and $t_2$. This evaluates the robustness of the decision manifold by verifying the stability of the global maximum:

\begin{equation}
\mathcal{S}_{\kappa}(t_1, t_2) = \kappa(s(z), t_1, t_2)
\end{equation}

\subsection{Automated Threshold Optimization Protocol}

To ensure that no framework is unfairly disadvantaged, we implement a Unified Optimization Protocol. For every scoring function $s(z)$, the decision boundaries are calibrated using a stratified grid-search on the validation set $V$:

\paragraph{Search Range:}

$t$ is constrained within the empirical percentiles of the validation scores:

\begin{equation}
t \in [P_{5}(s(V)), P_{95}(s(V))]
\end{equation}


The protocol strictly enforces the ordering of the boundaries:

\begin{equation}
t_1 < t_2
\end{equation}

The threshold pair $(t_1^*, t_2^*)$ is selected by maximizing the Cohen’s $\kappa$:
\begin{equation}
(t_1^*, t_2^*) = \arg\max_{t_1 < t_2} \kappa(s(V), y(V))
\end{equation}
\rthis{Once optimally calibrated across the cross-validation folds, these decision boundaries are strictly locked prior to evaluation on the 10,000-sample unseen hold-out set. This strict separation guarantees that our final performance metrics reflect true generalization capabilities without data leakage.}

\rthis{Collectively, these structural and numerical constraints inherently regularize the models against statistical outliers. Because the dataset relies on spectroscopically confirmed objects, severe instrumental errors are exceptionally rare, meaning anomalous scores typically represent physical edge cases rather than bad data. To prevent the mathematical search from contorting to fit these anomalies, we structurally enforce the parsimony constraint ($C \le 10$) and explicitly prune unstable motifs like nested exponentials. Furthermore, by strictly bounding the threshold grid-search within the 5th and 95th percentiles of the validation scores ($t \in [P_5, P_{95}]$), we numerically guarantee that extreme outlier predictions cannot disproportionately skew the final classification boundaries.}

\subsection{Symbolic Regression Algorithms}

Symbolic regression (SR) simultaneously discovers both the mathematical structure and numerical parameters of analytic expressions, offering a transparent alternative to traditional fixed-form regression. By representing candidate expressions as tree structures, SR navigates an expansive functional space to identify the most parsimonious mapping between input features and target scores.

In the context of SDSS DR17, this approach allows for the discovery of explicit decision functions, $s(z)$, that encode the astrophysical boundaries separating Galactic stars, galaxies, and high-redshift quasars. As detailed in Section~\ref{secpre}, each of the following four frameworks is evaluated under an identical \rthis{two-stage protocol: a five-fold stratified cross-validation to optimize decision thresholds ($t_1, t_2$), followed by a final evaluation on the unseen hold-out set}. This ensures that the comparison isolates each framework's intrinsic ability to capture the non-linear transition regions of the one-dimensional redshift manifold. We now discuss the different symbolic regression algorithms used in this work.
A comparative overview of the engines, optimization objectives, and inductive biases for these four symbolic frameworks is provided in Table \ref{tab:sr_clean}.

\subsubsection{{\tt PySR}: Evolutionary Island-Model Search}

We employ \textbf{{\tt PySR}}, a high-performance symbolic regression framework based on an island-model evolutionary algorithm. In this architecture, multiple semi-independent subpopulations (``islands'') evolve in parallel, with periodic migration of candidate expressions between islands. This strategy mitigates premature convergence and promotes structural diversity within the evolving symbolic population.

\paragraph*{Search Configuration and Operator Set}

The functional search space is defined by the binary operator set as follows:
\begin{equation}
\label{eq:2}
\mathcal{O}_{\text{binary}} = \{+, -, \times, \div\},
\end{equation}
and the unary operator set  below:
\begin{equation}
\label{eq:3}
\mathcal{O}_{\text{unary}} = \{\exp\}.
\end{equation}

To ensure numerical stability and physical consistency, protected operator variants are used for division, and nested exponential constraints are imposed to prevent unstable recursive growth. To prioritize interpretability and parsimony, the symbolic complexity is strictly bounded as shown below:

\begin{equation}
\label{eq:4}
\rthis{C(s)} \leq 10,
\end{equation}
where \rthis{$C(s)$ denotes the symbolic complexity, measured as the total number of nodes in the expression tree.}
The search objective minimizes the regularized $L^2$ loss, defined as follows: 
\begin{equation}
\label{eq:5}
\mathcal{L}_{\text{SR}} = \frac{1}{N} \sum_{i=1}^{N} \left( s(z_i) - y_i \right)^2,
\end{equation}
where $s(z)$ is the candidate symbolic function and $y_i$ denotes the target encoding, and $N$ represents the number of training samples. The evolutionary process is executed for 500 iterations in a representative 10,000-sample subset. This subset is partitioned into 5,000 training samples for functional evolution and 5,000 validation samples used for the selection of fitness-based models. The process is accelerated through {\tt Julia}-based multi-threading, enabling evaluation of thousands of candidate expressions per second.

\paragraph*{Evolutionary Strategy and Refinement}

The evolutionary process employs tournament selection combined with structural genetic operators, including subtree crossover, point mutation, and elitist reproduction. Algebraic simplification rules-specifically constant folding and identity elimination-are iteratively applied to maintain compact functional forms.

After structural discovery, numerical constants are refined via local {\tt BFGS} optimization:
\begin{equation}
\label{eq:6}
\theta^{*} = \arg\min_{\theta} \mathcal{L}_{\text{SR}}(s_{\theta}),
\end{equation}
where $\theta$ represents the continuous parameters embedded in the symbolic structure. This refinement improves numerical precision without modifying the discovered analytic form.

\paragraph{Model Selection and Thresholding}

\begin{figure}[htbp]
    \centering
    \includegraphics[width=1.0\columnwidth,height=0.22\textheight,keepaspectratio]{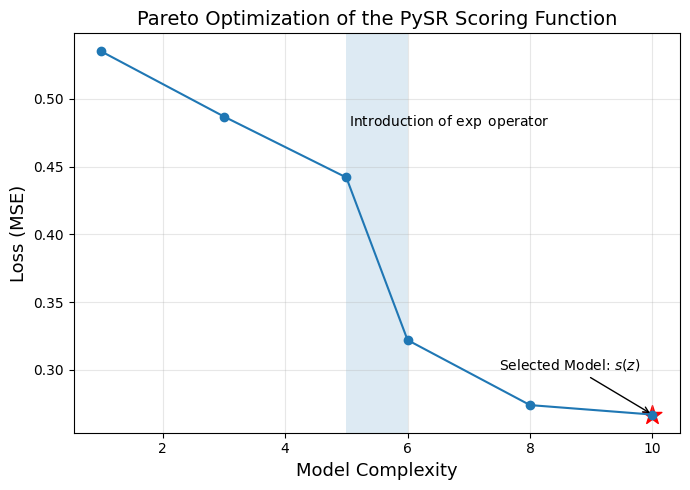}
    \caption{\rthis{Pareto-optimal frontier for the {\tt PySR} discovery phase. The plot illustrates the trade-off between functional complexity and $L_2$ loss. A significant performance gain is observed at $C=6$, corresponding to the introduction of the exponential ($\exp$) operator. The final scoring function $s(z)$ (indicated by the star) was selected at our predefined complexity boundary of $C=10$. While the algorithm could marginally reduce the loss by generating increasingly long and complex expressions, we deliberately truncated the selection at $C=10$ to prevent structural overfitting and guarantee that the final equation remains human-interpretable.}}
    \label{fig:pysr_landscape}
\end{figure}

\rthis{Although the internal search is guided by complexity-regularized $L_2$ loss, the final universal scoring function $s(z)$ is selected using a custom multi-objective validation metric. Because candidate equations output continuous scores, they are temporarily discretized using fixed provisional thresholds ($t_1=0.5, t_2=1.0$) during the discovery phase to approximate classification performance. The final representative scoring function is extracted from the Pareto-optimal frontier (visualized in Figure \ref{fig:pysr_landscape}) by ranking candidate expressions according to:}
\begin{equation}
\label{eq:7}
\text{Score} =
0.4 \cdot \text{Accuracy}
+ 0.4 \cdot \kappa
- 0.2 \cdot \frac{C(s)}{10}.
\end{equation}

\rthis{Here, Accuracy and $\kappa$ represent the standard classification accuracy and Cohen's Kappa, respectively, evaluated using the fixed provisional thresholds. The fixed thresholds and weighting factors (0.4, 0.4, 0.2) are adopted directly from \citet{Fabio25} to ensure a strict methodological comparison. As established in their work, assigning equal weight to Accuracy and $\kappa$ is a deliberate design choice to prevent the evolutionary search from collapsing into a majority-class predictor---a critical guardrail given the highly imbalanced nature of the SDSS dataset. Empirical sensitivity tests indicate that the resulting analytic forms are robust to minor weight perturbations ($\pm 0.1$); however, removing the 0.2 complexity penalty entirely routinely leads to structurally saturated, over-fitted expressions at the $C=10$ limit.}

Once the optimal analytic form $s(z)$ is selected, it is subjected to the unified evaluation protocol described in Section~3.4. Optimal decision boundaries $(t_1, t_2)$ are rigorously calibrated via grid search \rthis{during the 80,000-sample cross-validation phase} to explicitly maximize the final Cohen’s Kappa:

\begin{equation}
\label{eq:8}
(t_1, t_2) = \arg\max_{t_1 < t_2} \kappa.
\end{equation}

\rthis{This formal distinction clarifies our methodology: the $\kappa$ in Eq.~\ref{eq:7} is a provisional metric used strictly to guide internal structural discovery, whereas the $\kappa$ maximized in Eq.~\ref{eq:8} is the final, rigorously calibrated metric used to define global classification boundaries.}

\subsubsection{Exhaustive Symbolic Regression ({\tt ESR}) with MDL Selection}

To provide a deterministic baseline free from the stochastic variability inherent in evolutionary search, we implement Exhaustive Symbolic Regression ({\tt ESR}). Unlike heuristic methods, {\tt ESR} guarantees the evaluation of the complete mathematically valid candidate set within a predefined complexity limit, enabling principled and reproducible model selection.

\paragraph*{Search Space and Structural Filtering}

The candidate expressions are constructed from a basis library of binary operators
\begin{equation}
O_{\text{binary}} = \{+, -, \times, \div\},
\end{equation}
and the unary operator
\begin{equation}
O_{\text{unary}} = \{\exp\}.
\end{equation}

The terminal nodes consist of the spectroscopic redshift $z$ (as before) and learnable scalar constants $c_n$. To maintain computational tractability while ensuring deep functional exploration, we enforce a strict complexity bound
\begin{equation}
C(s) \leq 10.
\end{equation}

\rthis{Similar to the PySR framework, the complexity $C(s)$ denotes the number of nodes in the symbolic expression, ensuring a consistent measure of parsimony across all algorithms.}

To ensure astrophysical relevance and numerical stability, the framework applies three specific structural filters during the enumeration process:

\textbf{Redshift Requirement:} Expressions lacking the input variable $z$ are discarded to ensure the scoring function is dependent on the primary physical feature.

\textbf{Nested Exponential Filter:} Recursive exponential forms (e.g., $\exp(\exp(z))$) are prohibited to prevent mathematical instability and unphysical growth.

\textbf{Functional Constraints:} Exponential terms are restricted to valid physical forms, specifically targeting motifs like $\exp(c \cdot z)$, which characterize redshift-based density transitions.

\paragraph*{Parameter Optimization}

For expressions containing free constants, we employ an {\tt L-BFGS} optimization strategy to minimize the Mean Squared Error (MSE)~\citep{BFGS}. To ensure global convergence and prevent entrapment in local minima, the optimizer is initialized through a multi-start parallel routine. This process executes 30 independent optimizations per expression from randomized initial positions within a broad parameter space.

\paragraph*{Information-Theoretic Model Selection (MDL)}

\begin{figure}[htbp]
    \centering
    \includegraphics[width=1.0\columnwidth,height=0.22\textheight,keepaspectratio]{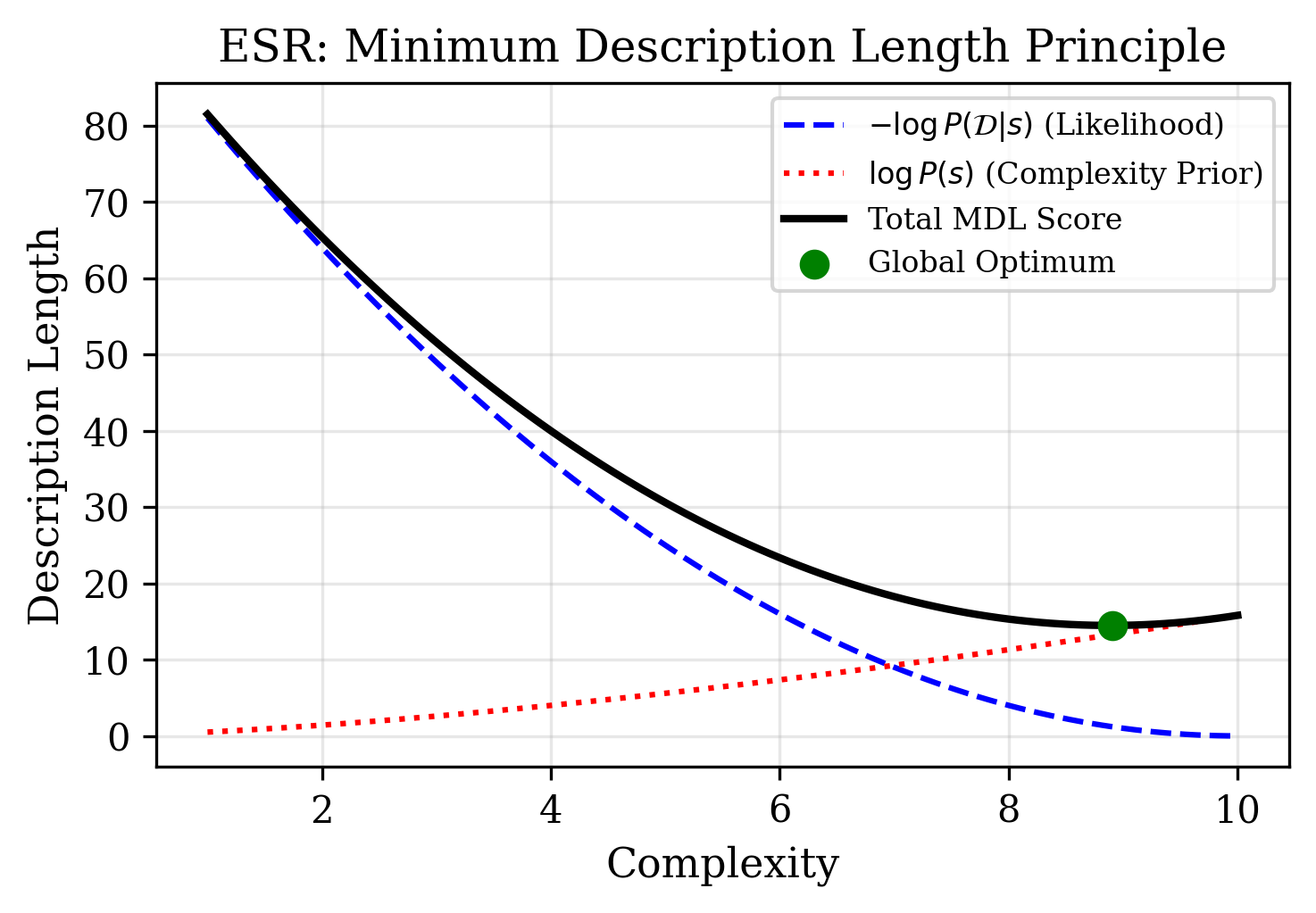}
    \caption{Model selection in {\tt ESR} via the Minimum Description Length (MDL) principle. The plot illustrates the information-theoretic competition between the negative log-likelihood (goodness of fit, Eq.~\ref{eq:negL}) and the complexity prior (parsimony penalty, Eq.~\ref{eq:Cs}). The final model is selected at the global minimum of the total description length (indicated by the green dot), ensuring the discovery of a non-overfitted, physically parsimonious scoring function as established in \citet{ESR}}
    \label{f6}
\end{figure}

Model selection is governed by the Minimum Description Length (MDL) principle. To ensure computational tractability during the exhaustive search, we implement the Bayesian Information Criterion (BIC) formulation~\citep{Krishak} of the MDL score. \rthis{As noted by \citet{ESR}, the BIC can be mathematically derived as a special case of the exact description length.} The MDL framework identifies the optimal model by balancing predictive fidelity against structural and parametric complexity. This information-theoretic competition is visualized in Figure \ref{f6}, where the final model is identified at the global minimum of the total description length. The MDL score for an expression $s$ is calculated as follows:

\begin{equation}
S_{\text{MDL}} =
\mathcal{L}_{\text{neg}}
+ C(s)\ln(N_{\text{ops}} + 1)
+ \frac{k}{2}\ln(N).
\label{eq:Cs}
\end{equation}
where,
$\mathcal{L}_{\text{neg}}$ denotes the negative log-likelihood, which can be calculated as follows:
\begin{equation}
\mathcal{L}_{\text{neg}} = 0.5\,N\ln(2\pi \cdot \text{MSE}) + 0.5\,N.
\label{eq:negL}
\end{equation}
The other terms in Eq.~\ref{eq:Cs} are defined as follows:
\begin{itemize}
\item $C(s)\ln(N_{\text{ops}} + 1)$: The structural complexity penalty based on the operator library size $N_{\text{ops}}$.

\item $\frac{k}{2}\ln(N)$: The penalty parameter for the $k$ optimized constants across $N$ data points.
\end{itemize}

\paragraph*{Evaluation Protocol}

Following the unified protocol established in Section~3.2, the top-ranked {\tt ESR} model per fold is used to generate continuous scores $s(z)$. Decision thresholds $t_1$ and $t_2$ are then calibrated \rthis{during the 80,000-sample cross-validation phase} via grid search to maximize the Cohen’s $\kappa$ coefficient, \rthis{before final evaluation on the hold-out set,} ensuring a rigorous benchmark against the evolutionary and reinforcement learning-based frameworks.

\subsubsection{Deep Reinforcement Learning Symbolic Regression ({\tt PhySO})}

We implement a reinforcement learning-based symbolic regression framework inspired by the Physical Symbolic Optimization ({\tt PhySO}) architecture. This method reformulates symbolic discovery as a sequential decision-making process, utilizing a deep neural network to navigate the functional search space while maintaining mathematical consistency.

\paragraph*{Expression Generation as a Sequential Process}

Symbolic expressions are generated in prefix notation using an autoregressive Long Short-Term Memory (LSTM) network. At each discrete time step $t$, the policy $\pi_\theta$ samples a token $a_t$ from a predefined vocabulary $V$:

\begin{equation}
V_{\text{op}} = \{+, -, \times, \div\}, 
\quad
V_{\exp} = \{\exp\}, 
\quad
V_{\text{term}} = \{z, c\}.
\label{eq:V}
\end{equation}
In Eq.~\ref{eq:V}, $z$ represents the spectroscopic redshift and $c$ denotes learnable scalar constants. The construction is formulated as a Markov Decision Process (MDP), where the RNN controller generates sequences up to a maximum length of 10 nodes, ensuring parsimonious mathematical structures.

\paragraph*{Structural Constraints and Filtering}

\begin{figure}[htbp]
    \centering
    \includegraphics[width=1.0\columnwidth,height=0.22\textheight,keepaspectratio]{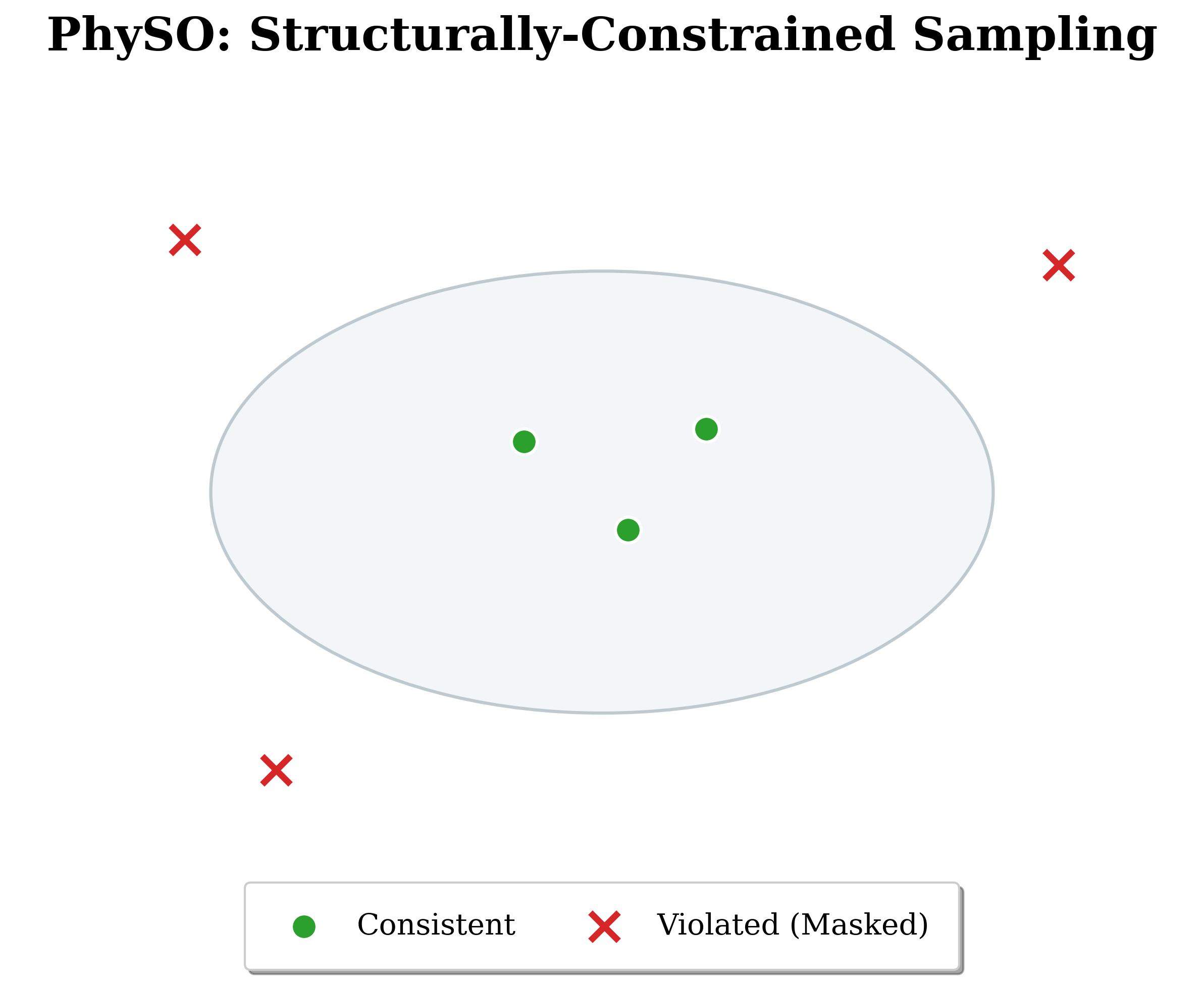}
    \caption{\textbf{Structurally-constrained sampling in {\tt PhySO}.} The reinforcement learning agent utilizes a masking function to eliminate candidate tokens that would violate mathematical or topological consistency (red crosses), such as invalid prefixes or nested exponentials. By restricting sampling to a valid functional manifold (shaded region), the framework efficiently identifies stable and mathematically plausible scoring motifs (green circles), adapting the original logic in \citet{Tenachi} for dimensionless feature spaces.}
    \label{fig:physo_search_space}
\end{figure}

To ensure the discovery of stable and \rthis{mathematically} plausible analytic forms, the framework applies a multi-layered filtering mechanism during the sampling process. As conceptually illustrated in Figure \ref{fig:physo_search_space}, the reinforcement learning agent employs a masking function to prune mathematically or \rthis{structurally} invalid paths in real-time:

\textbf{Redshift Requirement:} The generated expressions must contain the input variable $z$ to ensure functional dependence on the primary astrophysical feature.

\textbf{Nested Exponential Filter:} Recursive exponential forms (e.g., $\exp(\exp(z))$) are strictly prohibited to prevent mathematical instability and unphysical growth.

\textbf{Prefix Validity:} Sampled sequences are verified against operator parity rules; only mathematically sound expression trees are permitted for reward evaluation.

\rthis{While the original PhySO framework \citep{Tenachi} utilizes this masking mechanism to enforce physical unit constraints (e.g., preventing the addition of meters to seconds), our classification problem operates entirely on dimensionless variables (redshift $z$ and continuous scores). Therefore, we adapted the framework to bypass dimensional analysis, instead exclusively utilizing its prior-based masking to enforce strict mathematical and topological rules.} By aggressively masking out mathematically \rthis{or structurally} invalid paths, the neural network concentrates its search entirely within the space of consistent analytic forms.

\paragraph*{Policy Optimization and Reward Mechanism}

\rthis{To guide the structural search,} this approach emphasizes the top-performing fraction of sampled expressions-defined by a 90th percentile threshold-to encourage exploration of high-reward regions. The generator uses a softmax temperature of 1.15 to balance exploration and exploitation. The reward $R(s)$ for a generated expression $s(z)$ is derived from its predictive fidelity relative to the target ordinal labels $y$. This reward is bounded between $(0, 1]$ and is calculated as:
\begin{equation}
R(s) = \frac{1}{1 + \text{MSE}(s(z), y)}.
\end{equation}
The search is conducted on the initial 10,000-sample representative subset. It utilizes a batch size of 3,000 expressions per iteration over 200 generations, employing an exponential moving average (EMA) baseline with $\beta = 0.95$ to track search performance.

\paragraph*{Numerical Parameter Refinement}

For expressions containing free constants $c_n$, we utilize a high-performance {\tt L-BFGS} optimizer with a history size of 20 and Strong-Wolfe line search to refine numerical parameters. To maintain stability, our implementation utilizes protected operator variants, including clamped exponentials ($\exp(x)$ limited to $\pm 12.0$) and safe division (clamped at $\epsilon = 10^{-6}$), ensuring well-behaved gradients during optimization.

\subsubsection{Multi-View Symbolic Regression ({\tt MvSR})}

Multi-view symbolic regression addresses the non-stationarity of the SDSS redshift distribution by decomposing the discovery task into distinct functional regimes. This approach allows the framework to specialize in localized astrophysical structures while maintaining a globally consistent analytic template.

\begin{figure}[htbp]
    \centering
    \includegraphics[width=1.0\columnwidth,height=0.22\textheight,keepaspectratio]{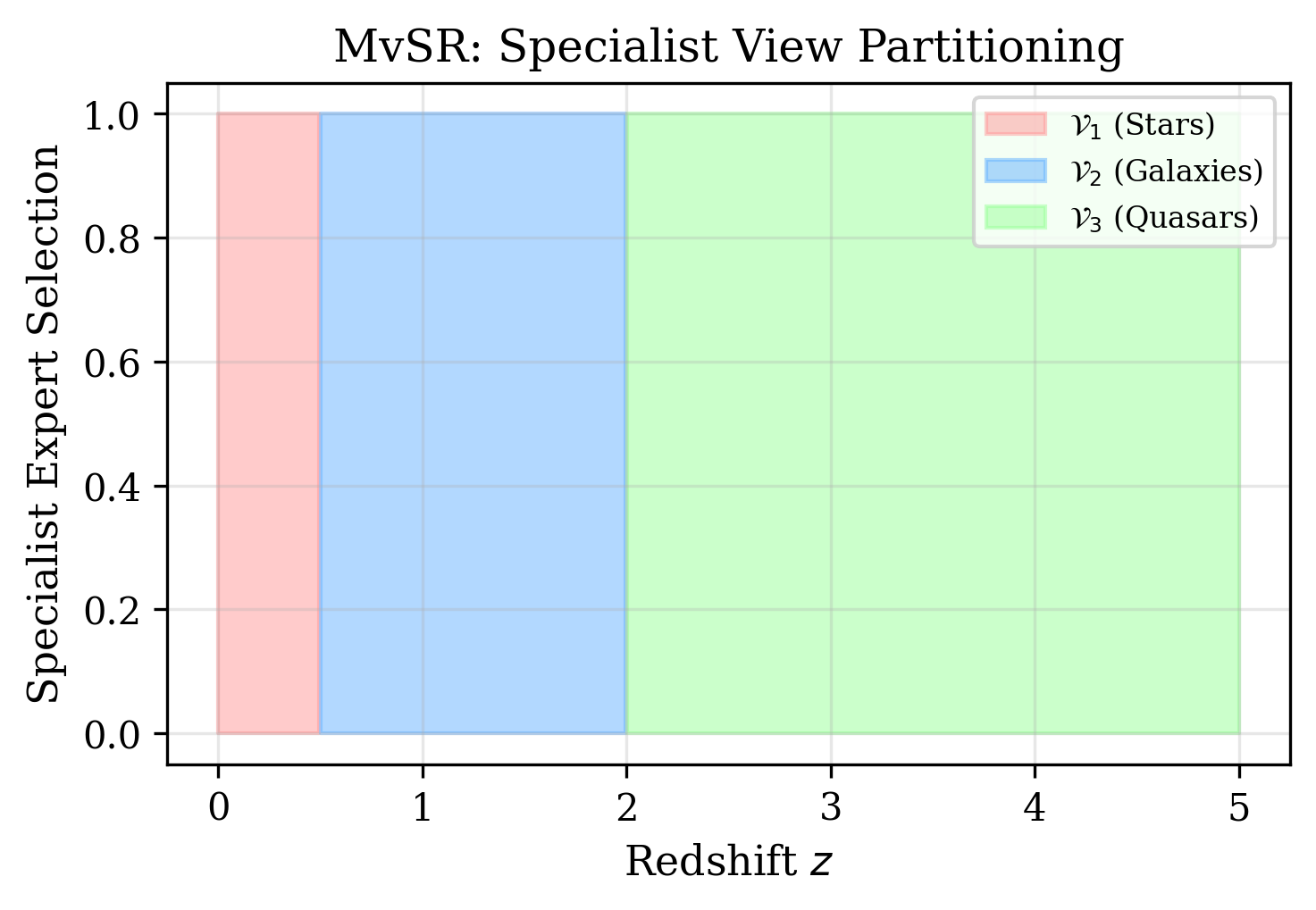}
    \caption{Specialist view partitioning in {\tt MvSR}. The redshift domain is decomposed into three contiguous, non-overlapping regimes: $\mathcal{V}_1$ (Stars), $\mathcal{V}_2$ (Galaxies), and $\mathcal{V}_3$ (Quasars). While stars ($\mathcal{V}_1$), galaxies ($\mathcal{V}_2$), and quasars ($\mathcal{V}_3$) dominate distinct contiguous regions, our {\tt MvSR} implementation captures this global variance by dividing the dataset into three equal-sized randomized segments. This multi-view decomposition enables the framework to adapt a single global symbolic template to these localized astrophysical structures by evolving independent numerical parameters for each randomized data cross-section, following the parametric multi-dataset integration logic in \citet{Russeil}.}
    \label{fig:mvsr_views}
\end{figure}

 \rthis{The underlying redshift manifold is characterized by three dominant object populations---stars, galaxies, and quasars---which occupy distinct transition regimes (conceptually visualized in Figure \ref{fig:mvsr_views}). To ensure the discovered analytic template is mathematically robust across these localized astrophysical structures, our implementation divides the dataset into three equal-sized randomized segments, ensuring each view captures a representative global population density.}Within each view, the data is further partitioned into 50\% training and 50\% validation samples to facilitate localized parameter refinement and objective assessment.

\paragraph*{Independent Evolution and Aggregated Fitness}

The search utilizes a genetic programming population of 1,000 individuals evolved over 35 generations with a basis library of
\begin{equation}
O_{\text{binary}} = \{+, -, \times, \div\},
\quad
O_{\text{unary}} = \{\exp\}.
\end{equation}

Unlike single-view methods, the fitness of a candidate expression $s(z)$ is determined by its Aggregated MSE, defined as the maximum error observed across all three views:

\begin{equation}
L_{\text{agg}} = \max\left( \text{MSE}_{\text{View1}}, \text{MSE}_{\text{View2}}, \text{MSE}_{\text{View3}} \right).
\end{equation}
This ``min-max'' objective ensures that the discovered functional form is robust across the entire redshift manifold and prevents the majority galaxy population from dominating the mathematical search.

\paragraph*{View-Specific Parameter Refinement}

While the symbolic structure (e.g., $s(z) = c_1 + c_2/(z + \exp(c_3 z))$) is globally shared, its numerical constants ($c_n$) are refined independently for each view. For every candidate expression, a GPU-accelerated {\tt L-BFGS} optimizer is used to find the optimal parameters $(\theta_1, \theta_2, \theta_3)$ that minimize the MSE within each respective data subset.

Following the unified protocol established in Section 3.2, the continuous scores generated by the best-performing multi-view model are used to calibrate decision thresholds $t_1$ and $t_2$. These thresholds are \rthis{optimized during the 80,000-sample cross-validation phase} via grid search to maximize the Cohen’s $\kappa$ coefficient, \rthis{prior to final evaluation on the hold-out set,} ensuring a rigorous benchmark against both single-view and exhaustive methods.

\begin{table*}[t]
\small
\centering
\caption{Comparison of Symbolic Regression Frameworks: Engine, Objective, Optimization, and Scientific Inductive Bias}
\label{tab:sr_clean}
\renewcommand{\arraystretch}{1.15}
\begin{tabularx}{\textwidth}{@{}l l X X@{}}
\toprule
\textbf{Method} & \textbf{Engine} & \textbf{Objective \& Optimization} & \textbf{Inductive Bias} \\
\midrule

\textbf{{\tt PySR}} 
& Island-model GP (Julia) 
& $\mathcal{L}=\frac{1}{n}\sum (y_i-s(z_i))^2+\alpha C$; local {\tt L-BFGS} constant refinement per expression. 
& Functional survival through migratory evolutionary competition across independent populations. \\

\addlinespace

\textbf{{\tt ESR}} 
& GPU exhaustive enumeration 
& $S_{MDL}=\mathcal{L}_{neg}+\frac{k}{2}\ln(n)+C\ln(N_{ops}+1)$; {\tt L-BFGS} with 30 multi-start initializations. 
& Occam’s Razor; prioritizes the simplest globally valid mathematical structure. \\

\addlinespace

\textbf{\tt PhySO} 
& Autoregressive LSTM sampling 
& Reward maximization ($R \geq P_{90}$); integrated {\tt L-BFGS} tuning during generation. 
& Dimensional stability via prior-based masking of invalid search paths. \\

\addlinespace

\textbf{{\tt MvSR}} 
& View-decomposed genetic programming 
& $L_{\text{agg}} = \max(\text{MSE}_{V1} , \text{MSE}_{V2} , \text{MSE}_{V3} )$; independent {\tt L-BFGS} refinement per view. 
& Robustness verification across randomized data partitions. \\

\bottomrule
\end{tabularx}
\end{table*}

\subsection{Machine Learning Benchmarks}

To establish a formal performance ceiling for the one-dimensional redshift manifold, 
we implement three distinct machine learning paradigms, following the theoretical foundations of pattern recognition and the connectionist framework established in \citet{Bishop2006}. The inclusion of these high-capacity benchmarks-specifically Random Forests, SVMs, and MLPs-is essential to define the Information-Theoretic Ceiling of the spectroscopic redshift distribution. As noted by \citet{pysr}, symbolic regression is an NP-hard optimization problem where the functional search space is uncountably infinite. Therefore, we utilize these non-interpretable paradigms to establish a definitive performance upper bound. These models serve as a control group to quantify the trade-off between black-box accuracy and symbolic transparency, allowing us to determine if the discovered expressions are physically sufficient or if they suffer from a 'complexity deficit' compared to overparameterized architectures. All models are restricted to the feature space 
$X = \{z\}$ to ensure a direct comparative basis with the symbolic frameworks. The paradigms, optimization strategies, and underlying inductive biases for these non-interpretable benchmarks are summarized in Table \ref{tab:ml_baselines}.

\subsubsection{Random Forest (RF)}

The Random Forest serves as a non-parametric ensemble baseline. It approximates 
the classification function $f(z)$ through an ensemble of $B \in \{100, 200\}$ randomized decision trees. Each tree $T_b$ partitions the redshift space into disjoint 
regions by identifying optimal split points $s$ that minimize the Gini impurity:
\begin{equation}
G = 1 - \sum_{k=1}^{K} p_k^2 , 
\end{equation}
where $p_k$ denotes the fraction of samples belonging to class $k$ within a 
given node. The final classification is determined via majority voting:
\begin{equation}
\hat{C}_{\text{RF}}(z) 
= \operatorname{mode} \left\{ T_1(z), T_2(z), \dots, T_B(z) \right\}
\end{equation}
To prevent structural over-complexity, the trees are constrained to a maximum 
depth of 10, allowing the ensemble to model high-frequency population transitions 
while maintaining stability.

\subsubsection{Support Vector Machine (SVM)}

The SVM evaluates the separability of the manifold by projecting the one-dimensional 
input into a high-dimensional Hilbert space using a Radial Basis Function (RBF) kernel:

\begin{equation}
K(z, z') = \exp\left( -\gamma \| z - z' \|^2 \right)
\end{equation}

The classifier identifies optimal separating hyperplanes by maximizing the 
margin between astrophysical classes. With an RBF kernel and a regularization parameter optimized across $C \in \{1, 10, 100\}$, the SVM constructs smooth, non-linear decision boundaries, 
serving as a benchmark for global margin maximization without explicit 
functional-form constraints.

\subsubsection{Multi-Layer Perceptron (MLP)}

The MLP represents the connectionist approach, where the mapping 
$z \rightarrow \text{Class}$ is learned through hidden layers. 
We implement a $(32,16)$ architecture following feature standardization to zero mean and unit variance, where each layer computes

\begin{equation}
h^{(l)} = \sigma \left( W^{(l)} h^{(l-1)} + b^{(l)} \right),
\end{equation}
with $\sigma(x) = \max(0, x)$ denoting the ReLU activation function.
This model quantifies the performance of a high-capacity neural network relying 
on thousands of learned parameters, contrasting with the compact symbolic nodes 
explored in this work.

\subsubsection{Stratified Holdout Generalization Protocol}

\rthis{To ensure a strictly fair comparison with the symbolic regression frameworks, we subject these high-capacity baselines to the exact same rigorous two-stage evaluation protocol. Each machine learning model is trained and hyperparameter-optimized (via an automated GridSearchCV protocol) using 5-fold cross-validation exclusively on the 80,000-sample optimization set.}

\rthis{Subsequently, the final models are evaluated on the identical 10,000-sample unseen hold-out test set. This protocol isolates \emph{generalization stability}, ensuring that the reported performance reflects genuine capture of the astrophysical manifold rather than localized over-fitting, and directly defines the maximum extractable information from the one-dimensional redshift feature space.}

\begin{table*}[t]
\small
\centering
\caption{Comparison of Machine Learning Baselines: Paradigm, Objective, Optimization, and Inductive Bias}
\label{tab:ml_baselines}
\renewcommand{\arraystretch}{1.15}
\begin{tabularx}{\textwidth}{@{}l l X X@{}}
\toprule
\textbf{Model} & \textbf{Paradigm} & \textbf{Objective \& Optimization} & \textbf{Inductive Bias} \\
\midrule

\textbf{Random Forest (RF)}
& Ensemble learning
& Minimization of Gini impurity via recursive axis-aligned partitioning; greedy heuristic search for locally optimal split points.
& Ensemble averaging; assumes that the consensus of decorrelated decision trees smooths decision boundaries along population density gradients. \\

\addlinespace

\textbf{Support Vector Machine (SVM)}
& Kernel-based margin maximization
& Maximization of the geometric margin under soft constraints; convex quadratic programming with RBF kernel transformation.
& Maximum-margin principle; favors globally smooth, high-separation decision boundaries in transformed feature space. \\

\addlinespace

\textbf{Multi-Layer Perceptron (MLP)}
& Feed-forward neural network
& Cross-entropy loss minimization via stochastic gradient descent (backpropagation) across layered nonlinear transformations.
& Distributed representation learning; assumes hierarchical nonlinear feature composition captures latent manifold structure. \\

\bottomrule
\end{tabularx}
\end{table*}

\section{Results and Analysis}
\label{sec:results}

We present a comprehensive evaluation of the symbolic regression (SR) frameworks, detailing the optimized analytic expressions, classification performance across the redshift manifold, and physical failure modes. \rthis{As established in Section~3.2, all results are derived using a rigorous two-stage protocol: an 80,000-sample 5-fold cross-validation threshold optimization phase, followed by final evaluation on a 10,000-sample unseen hold-out test set.}

\subsection{Framework-Specific Results}

For each symbolic framework, we provide a diagnostic suite consisting of the discovered scoring function, average confusion matrices, threshold sensitivity analysis, and class-specific recall profiles. \rthis{A consolidated overview of the performance metrics during the cross-validation optimization phase is provided in Table \ref{tab:sr_results_summary}, while the final generalization metrics on the unseen data are detailed in Table \ref{tab:sr_holdout_summary}.}

\subsubsection{{\tt PySR}: Evolutionary Island-Model Discovery}

\paragraph*{Discovered Expression}

The representative scoring function identified by {\tt PySR} captures non-linear 
population transitions using linear redshift scaling and a sharp exponential decay:

\begin{equation}
s(z) = 0.8395\,z + e^{-78.28 z} - 0.1141
\label{eq:pysr_score}
\end{equation}

\paragraph*{Classification Performance and Stability}

\begin{figure}[htbp]
    \centering
    \includegraphics[width=\linewidth,height=0.6\textheight,keepaspectratio]{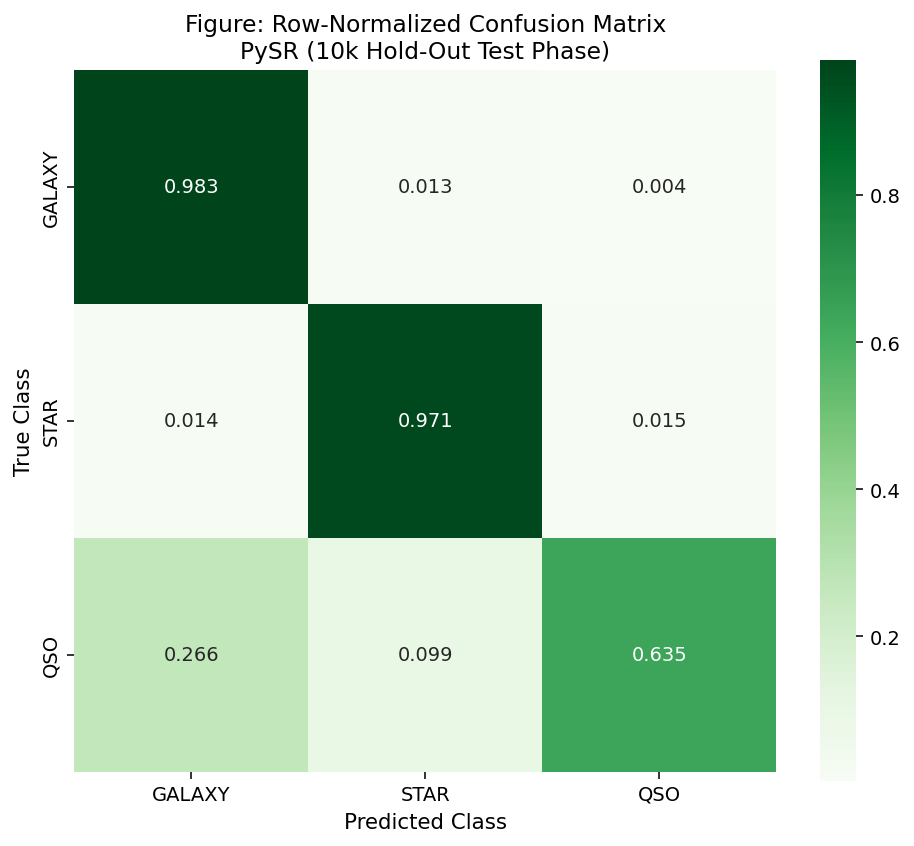}
    \caption{\rthis{Row-Normalized Confusion Matrix - {\tt PySR} (10k Hold-Out Test Phase)}. 
    The heatmap shows the fraction of actual classes correctly predicted 
    as GALAXY (0.983), STAR \rthis{(0.971)}, and QSO \rthis{(0.635)}.}
    \label{fig:f9}
\end{figure}

{\tt PySR} achieves a \rthis{cross-validation mean Cohen’s $\kappa$ of $0.8536 \pm 0.0032$ (accuracy $91.99 \pm 0.0017\%$), and a final hold-out test $\kappa$ of $0.8436$ (accuracy $91.47\%$).} 
The row-normalized confusion matrix \rthis{for the hold-out set}, shown in Figure \ref{fig:f9},
demonstrates exceptional fidelity for GALAXY (0.983) and 
STAR \rthis{(0.971)} populations.

However, a non-negligible fraction of QSO samples \rthis{(0.266)} is 
misclassified as galaxies, reflecting spectral overlap in a 
one-dimensional redshift feature space. This results in a 
class-specific $F1_{\text{QSO}}$ of \rthis{$0.7636$}. 
The framework exhibits \rthis{high stability} in its decision 
boundaries, with a threshold standard deviation of 
$\sigma_{t1} = \rthis{0.005}$.

\paragraph*{Error and Physical Failure Analysis}

\begin{figure}[htbp]
    \centering
    \includegraphics[width=\linewidth,height=0.6\textheight,keepaspectratio]{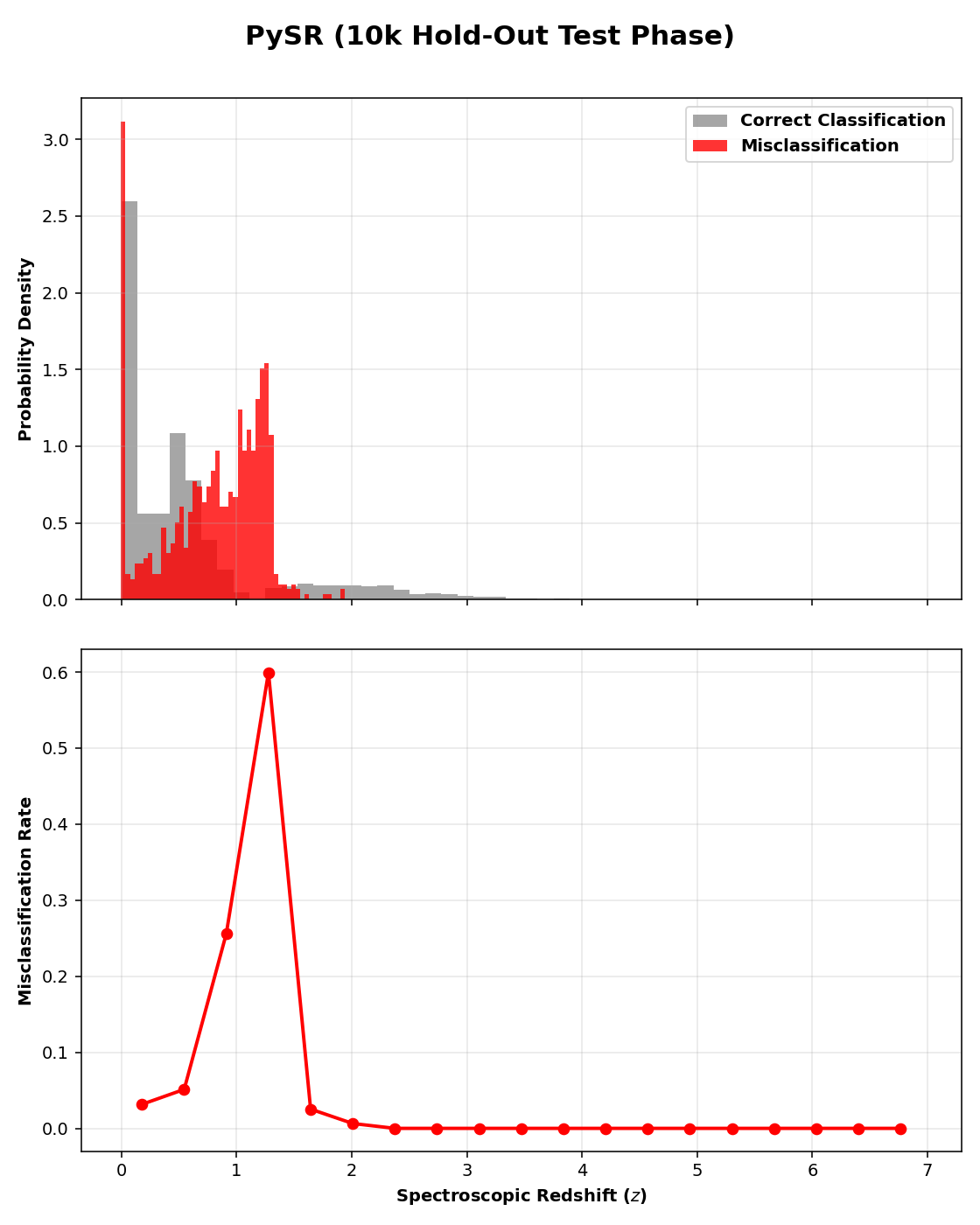}
    \caption{{\tt PySR} Error Analysis \rthis{(10k Hold-Out Test Phase)}. Top: Error density distribution showing correctly classified objects (gray) versus misclassifications (red) across redshift. Bottom: The Error rate vs. redshift profile, revealing a sharp failure spike peaking \rthis{at approximately 0.58 near $z \approx 1.25$}.}
    \label{fig:f11}
\end{figure}

Analysis of the error distribution, visualized in Figure \ref{fig:f11}, pinpoints the physical boundaries where the analytic form reaches its limit.

Misclassifications are nearly absent at $z \approx 0$ 
(stellar locus) but become prominent between 
$z \approx 0.8$ and $z \approx 1.5$, where the analytic form 
struggles to separate overlapping galaxy–QSO populations.

\paragraph*{Threshold Robustness and Model Stability}

\begin{figure}[htbp]
    \centering
    \includegraphics[width=\linewidth]{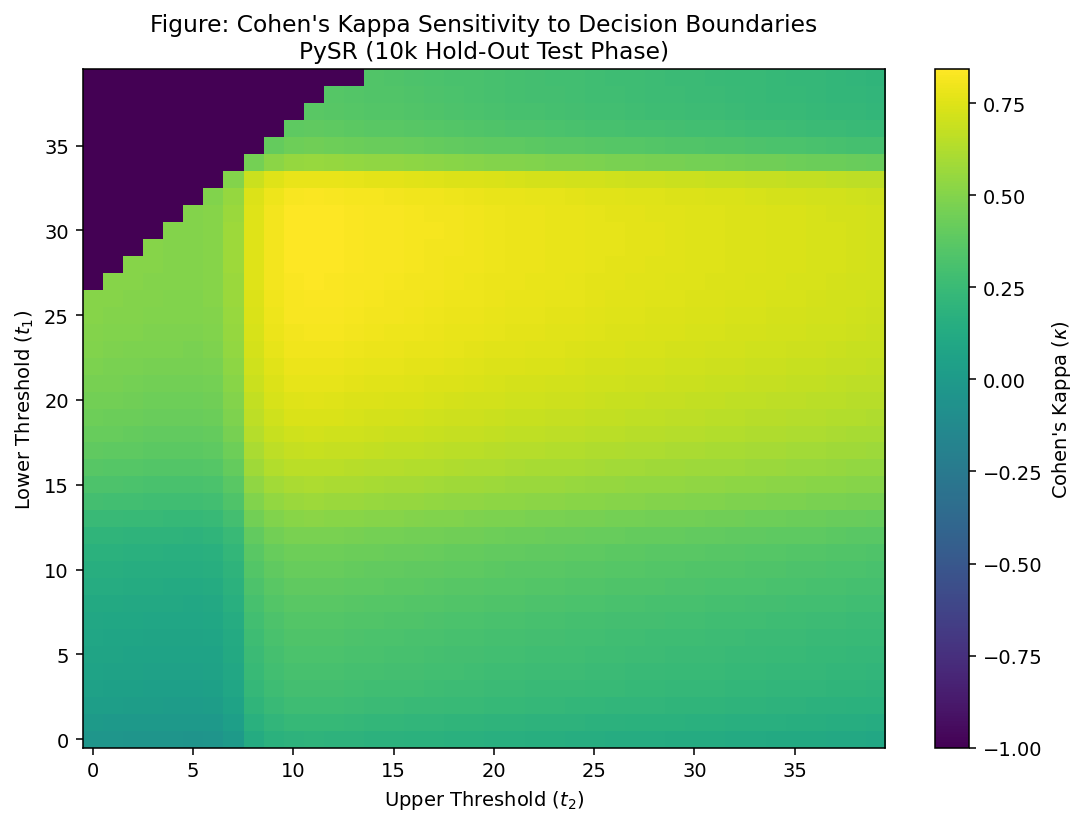}
    \caption{Kappa Sensitivity Heatmap - {\tt PySR} \rthis{(10k Hold-Out Test Phase)}. 
    A 2D grid visualizing Cohen’s $\kappa$ as a function of 
    decision thresholds $t_1$ and $t_2$. 
    A broad high-value plateau indicates stable class separation.}
    \label{fig:f10}
\end{figure}

The robustness of the discovered decision manifold is illustrated 
through the Kappa sensitivity landscape in Figure \ref{fig:f10}.

A broad high-value plateau is visible across a wide range 
of $t_1$ and $t_2$ combinations, indicating that $s(z)$ is not 
overly sensitive to minor threshold perturbations. 
The optimized thresholds remain \rthis{highly} consistent across \rthis{the calibration} folds:

\begin{equation}
t_1 = 0.811 \pm 0.005,
\qquad
t_2 = 0.981 \pm 0.009.
\end{equation}


\paragraph*{Recall Profiles and Physical Transition Dynamics}

\begin{figure}[htbp]
    \centering
    \includegraphics[width=\linewidth,height=0.6\textheight,keepaspectratio]{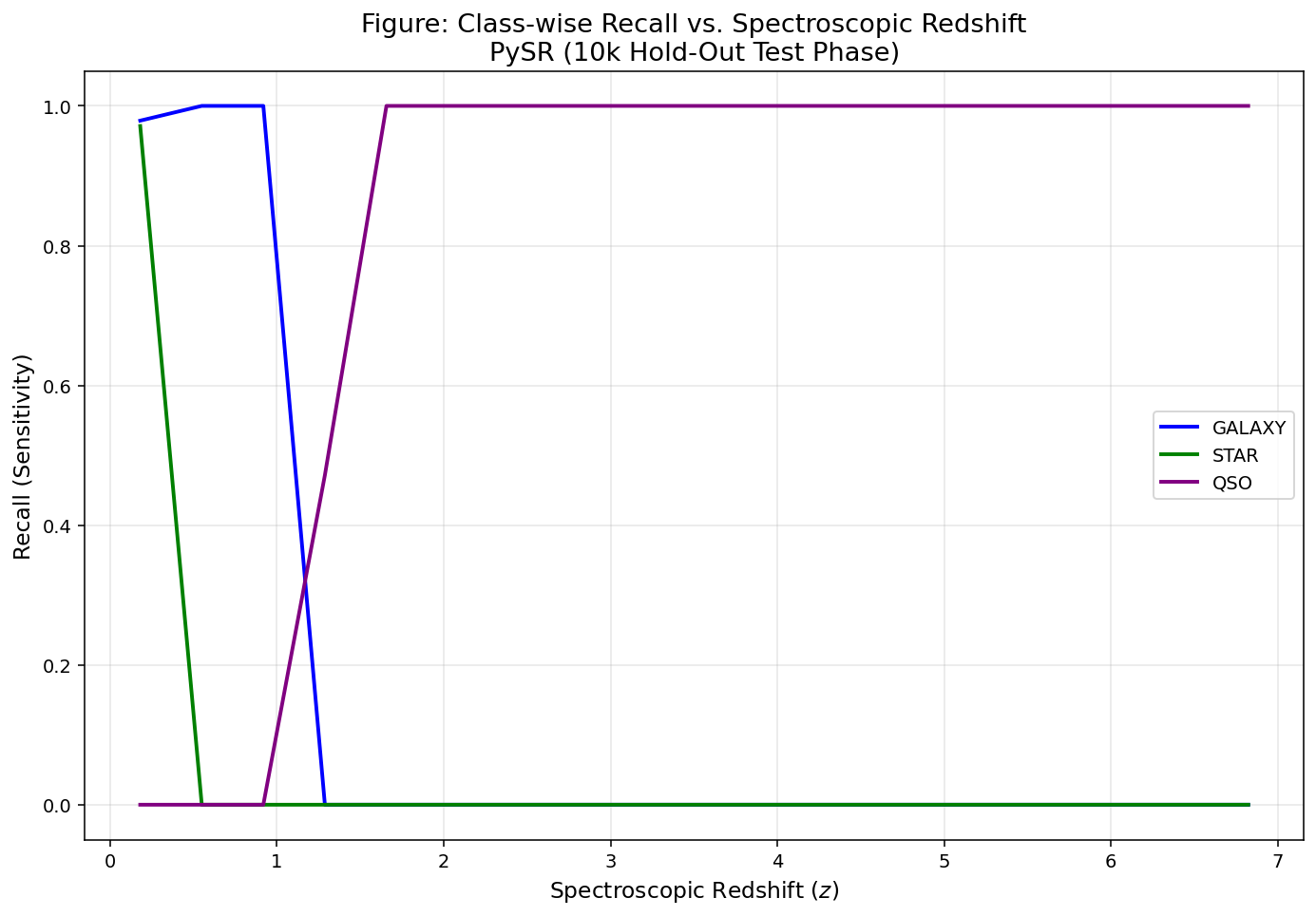}
    \caption{Recall vs. Redshift - Physical Failure Analysis ({\tt PySR} \rthis{10k Hold-Out Test Phase}). 
    Class-wise recall across redshift \rthis{illustrating the population transition boundaries}.}
    \label{fig:f12}
\end{figure}

The Recall vs. Redshift profile (Figure \ref{fig:f12}) summarizes the physical 
transition behavior of the classifier \rthis{on the unseen data}.

STAR recall remains near unity for $z < 0.5$ before sharply 
declining as the GALAXY regime emerges. GALAXY recall 
stays near 1.0 until $z \approx 0.9$, after which it drops 
rapidly as QSO recall rises to unity.

\subsubsection{{\tt ESR}: Exhaustive MDL-Based Selection}

\paragraph*{Discovered Expression}

The deterministic search provided by {\tt ESR} favors a parsimonious quadratic 
form with a dampened exponential term, ensuring a structurally stable 
functional mapping:

\begin{equation}
s(z) = 0.7 + z^2 + e^{-11.88 z}
\label{eq:esr_score}
\end{equation}

\paragraph*{Classification Performance and Stability}

{\tt ESR} demonstrates superior quasar identification relative to the 
evolutionary baseline, yielding a \rthis{hold-out test} $F1_{\text{QSO}}$ of 
\rthis{$0.8074$}. The framework achieves a \rthis{cross-validation mean Cohen’s 
$\kappa$ of $0.8664 \pm 0.0034$ (accuracy $92.59 \pm 0.0021\%$), and a final hold-out test $\kappa$ of $0.8638$ (accuracy $92.46\%$).}

\begin{figure}[htbp]
    \centering
    \includegraphics[width=\linewidth,height=0.6\textheight,keepaspectratio]{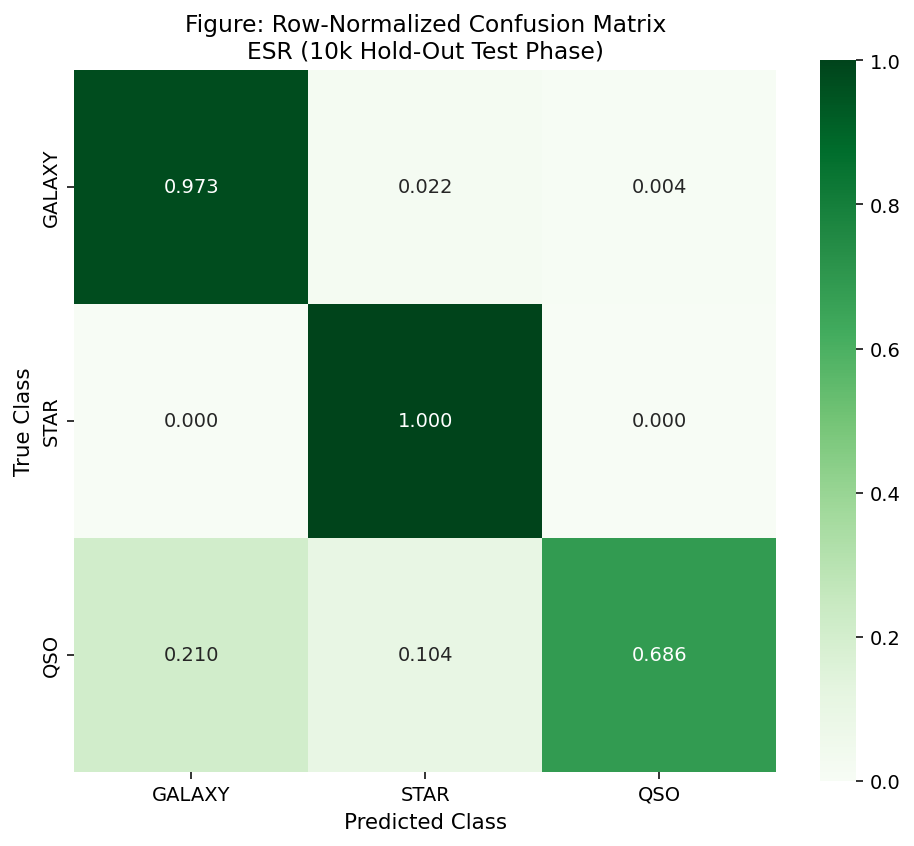}
    \caption{\rthis{Row-Normalized Confusion Matrix - {\tt ESR} (10k Hold-Out Test Phase)}. 
    The heatmap shows the fraction of actual classes correctly predicted 
    as GALAXY \rthis{(0.973)}, STAR \rthis{(1.000)}, and QSO \rthis{(0.686)}.}
    \label{fig:f13}
\end{figure}

The row-normalized confusion matrix \rthis{for the hold-out set}, shown in Figure \ref{fig:f13}, confirms robust 
performance across all classes, with particularly high fidelity 
for STAR \rthis{(1.000)} and GALAXY \rthis{(0.973)} populations.

The optimized decision thresholds remain \rthis{highly} consistent across \rthis{the calibration} folds:
\begin{equation}
t_1 = 1.654 \pm 0.027,
\qquad
t_2 = 2.184 \pm 0.033.
\end{equation}


\paragraph*{Error and Physical Failure Analysis}

Analysis of the error distribution reveals that misclassifications 
are concentrated within the range $z \approx 0.8$ to $z \approx 1.5$.

\begin{figure}[htbp]
    \centering
    \includegraphics[width=\linewidth,height=0.6\textheight,keepaspectratio]{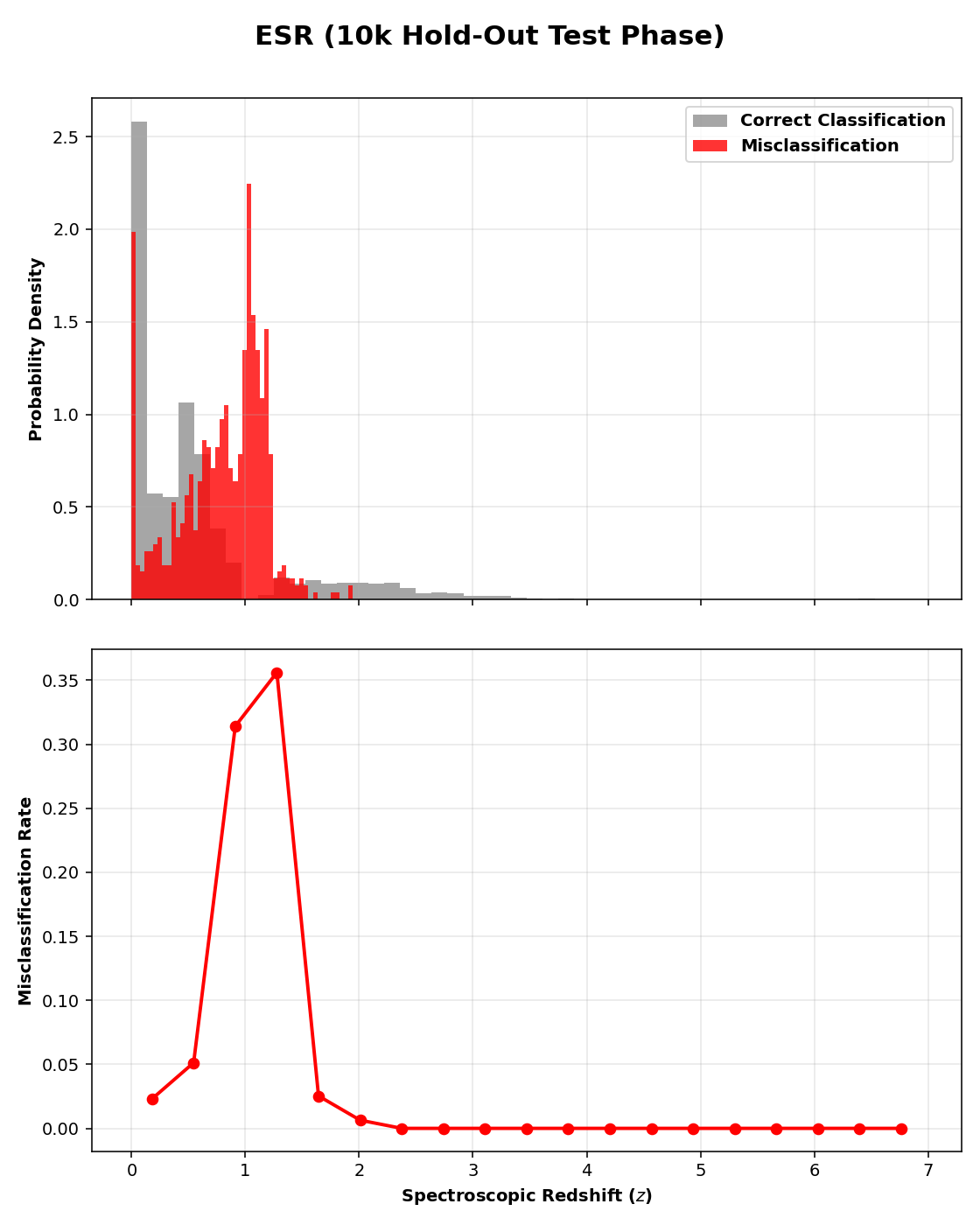}
    \caption{{\tt ESR} Error Analysis \rthis{(10k Hold-Out Test Phase)}. Top: Error density distribution showing correctly classified objects (gray) versus misclassifications (red) across redshift. Bottom: Error rate vs. redshift profile indicating a peak failure rate of approximately \rthis{0.38} near $z \approx 1.25$.}
    \label{fig:f15}
\end{figure}

The Error Rate vs. Redshift profile (Figure \ref{fig:f15}) demonstrates a peak failure 
rate of approximately \rthis{0.38} near $z \approx 1.25$, significantly 
more suppressed than the spikes observed in heuristic search methods.

\paragraph*{Threshold Robustness and Model Stability}

\begin{figure}[htbp]
    \centering
    \includegraphics[width=\linewidth,height=0.6\textheight,keepaspectratio]{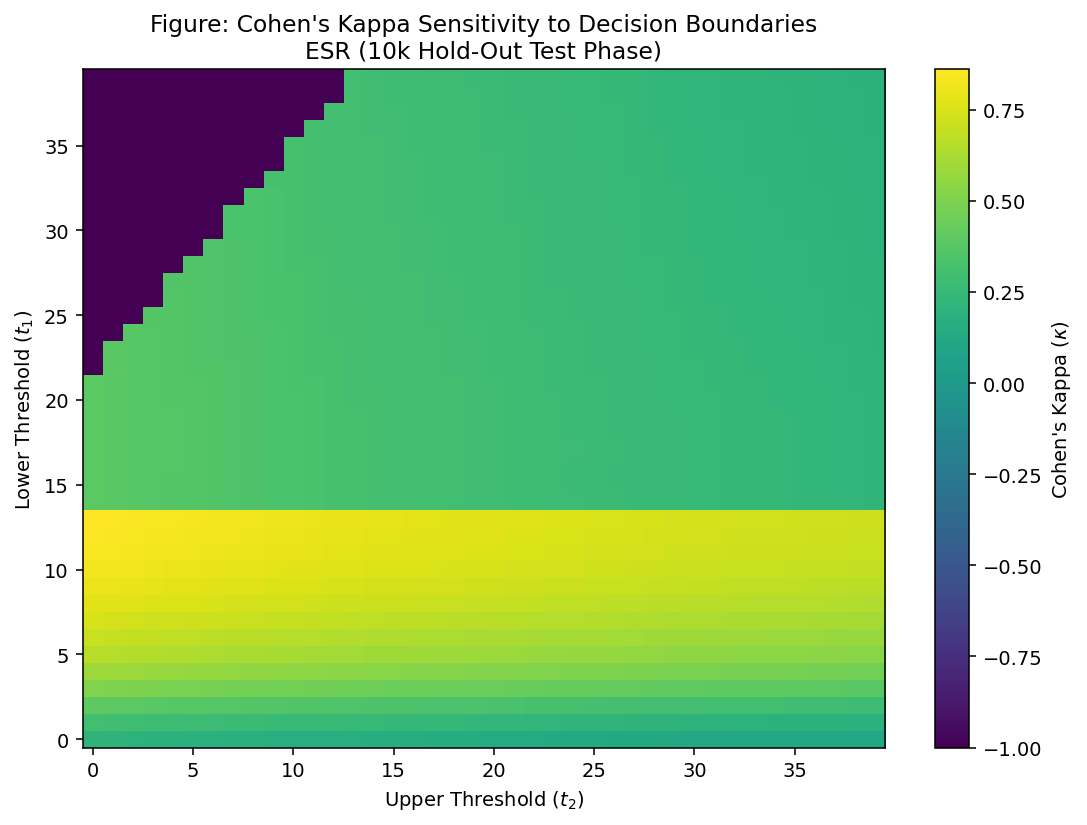}
    \caption{Kappa Sensitivity Heatmap - {\tt ESR} \rthis{(10k Hold-Out Test Phase)}. A 2D grid visualizing 
    Cohen’s $\kappa$ as a function of decision thresholds $t_1$ and 
    $t_2$, highlighting an expansive optimal plateau.}
    \label{fig:f14}
\end{figure}

The robustness of the MDL-selected decision manifold is illustrated 
through the Kappa sensitivity landscape in Figure \ref{fig:f14}.

The extensive high-value plateau indicates that the analytic form 
is less sensitive to minor threshold shifts, suggesting that the 
MDL-based selection successfully identifies a physically robust 
scoring manifold that generalizes well across different data samples.
\paragraph*{Recall Profiles and Physical Transition Dynamics}

\begin{figure}[htbp]
    \centering
    \includegraphics[width=\linewidth,height=0.6\textheight,keepaspectratio]{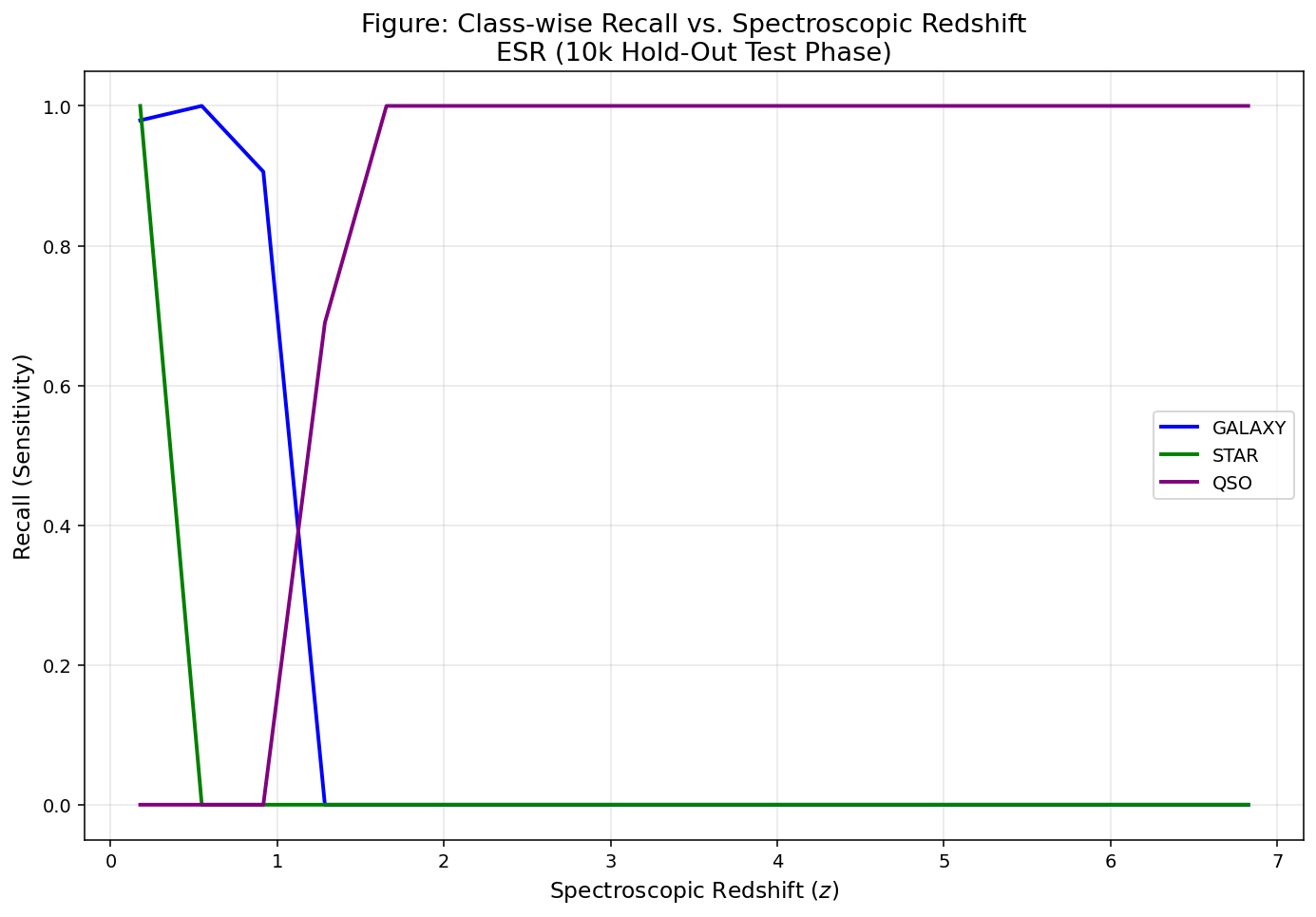}
    \caption{Recall vs. Redshift - Physical Failure Analysis ({\tt ESR} \rthis{10k Hold-Out Test Phase}). 
    Line plot displaying recall for each class across the redshift 
    range, \rthis{illustrating the population transition boundaries}.}
    \label{fig:f16}
\end{figure}

The curve for recall as a function of redshift (Figure \ref{fig:f16}) highlights the framework's 
reliable recovery of the stellar population ($z < 0.5$).

Similar to other models, a characteristic crossover exists between 
$z \approx 1.0$ and $z \approx 1.5$, where GALAXY recall falls 
as QSO recall rises, defining the physical boundary of 
one-dimensional redshift separation.


\subsubsection{{\tt PhySO}: Deep Reinforcement Learning Optimization}

\paragraph*{Discovered Expression}

The reinforcement learning agent discovers a nested functional motif 
that precisely maps redshift transitions by identifying stable, 
Gaussian-like structures:

\begin{equation}
s(z) = 2.4027 - e^{\,0.5786 - z^2 + z}
\label{eq:physo_score}
\end{equation}


\paragraph*{Classification Performance and Stability}

{\tt PhySO} exhibits exceptional parametric stability, achieving the lowest 
threshold standard deviation in the study:

\begin{equation}
t_1 = 0.587 \pm 0.001,
\qquad
t_2 = 0.908 \pm 0.002.
\end{equation}

The framework yields a \rthis{cross-validation mean Cohen’s $\kappa$ of 
$0.8753 \pm 0.0014$ (accuracy $93.08 \pm 0.0007\%$), and a final hold-out test $\kappa$ of $0.8670$ (accuracy $92.63\%$).}

\begin{figure}[htbp]
    \centering
    \includegraphics[width=\linewidth,height=0.6\textheight,keepaspectratio]{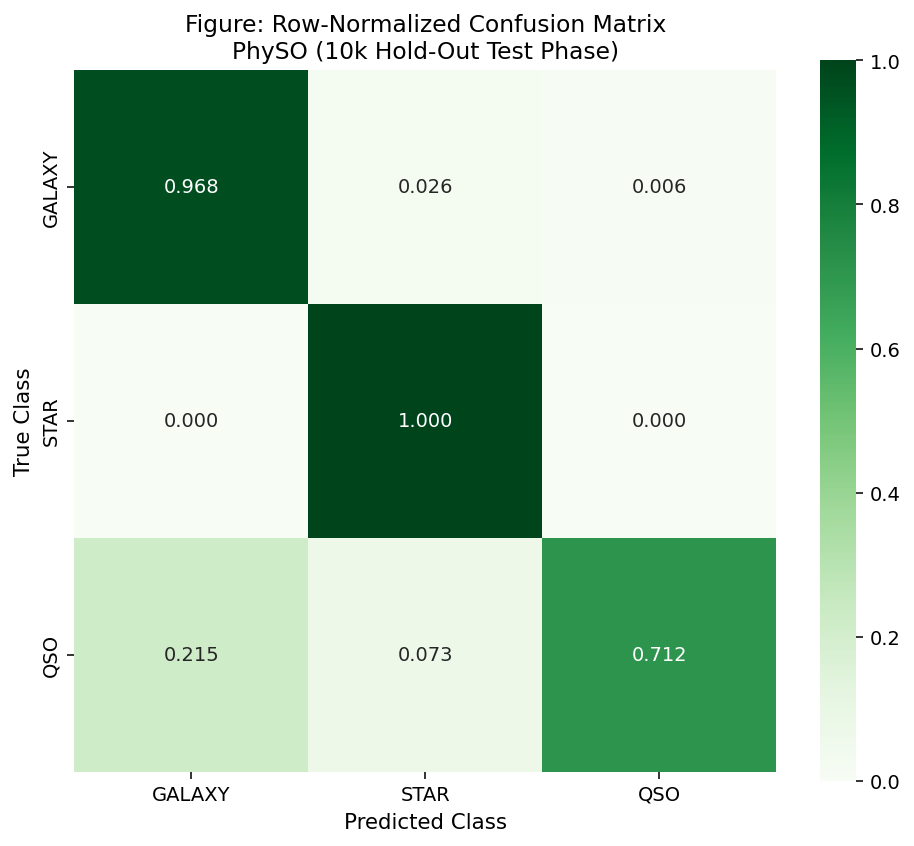}
    \caption{\rthis{Row-Normalized Confusion Matrix - {\tt PhySO} (10k Hold-Out Test Phase)}. 
    The heatmap illustrates the fraction of actual classes correctly 
    predicted, featuring perfect STAR identification (1.000) and 
    high GALAXY \rthis{(0.968)} accuracy.}
    \label{fig:f17}
\end{figure}

The row-normalized confusion matrix \rthis{for the hold-out set}, illustrated in Figure \ref{fig:f17}, demonstrates 
nearly perfect identification of the STAR population (1.000) 
and high fidelity for GALAXY \rthis{(0.968)}, while maintaining a 
robust QSO identification rate of \rthis{0.712}.

\paragraph*{Error and Physical Failure Analysis}

Analysis of the error distribution indicates that 
misclassifications are almost entirely absent at $z \approx 0$, but concentrate between $z \approx 0.8$ and $z \approx 1.5$.

\begin{figure}[htbp]
    \centering
    \includegraphics[width=\linewidth,height=0.6\textheight,keepaspectratio]{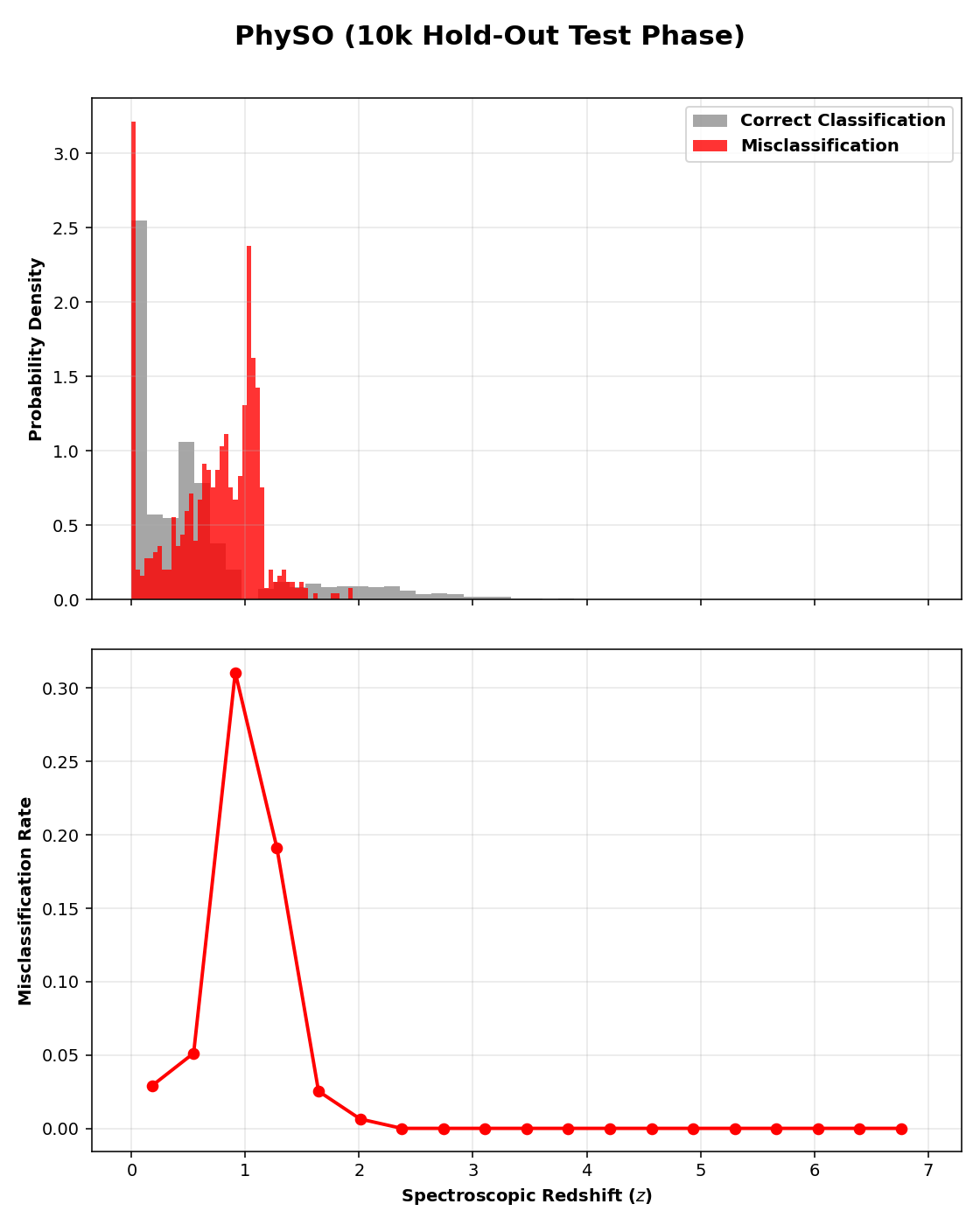}
    \caption{{\tt PhySO} Error Analysis \rthis{(10k Hold-Out Test Phase)}. Top: Error density distribution comparing correctly classified objects (gray) versus misclassifications (red) across redshift. Bottom: Error rate vs. redshift profile showing a peak error rate of approximately \rthis{0.38} near $z \approx 1.0$.}
    \label{fig:f19}
\end{figure}

The Error Rate vs. Redshift profile, visualized in Figure \ref{fig:f19}, reveals a suppressed 
peak failure rate of approximately \rthis{0.38} near 
$z \approx 1.0$, characterizing a smoother transition 
manifold than evolutionary baselines.

\paragraph*{Threshold Robustness and Model Stability}

\begin{figure}[htbp]
    \centering
    \includegraphics[width=\linewidth,height=0.6\textheight,keepaspectratio]{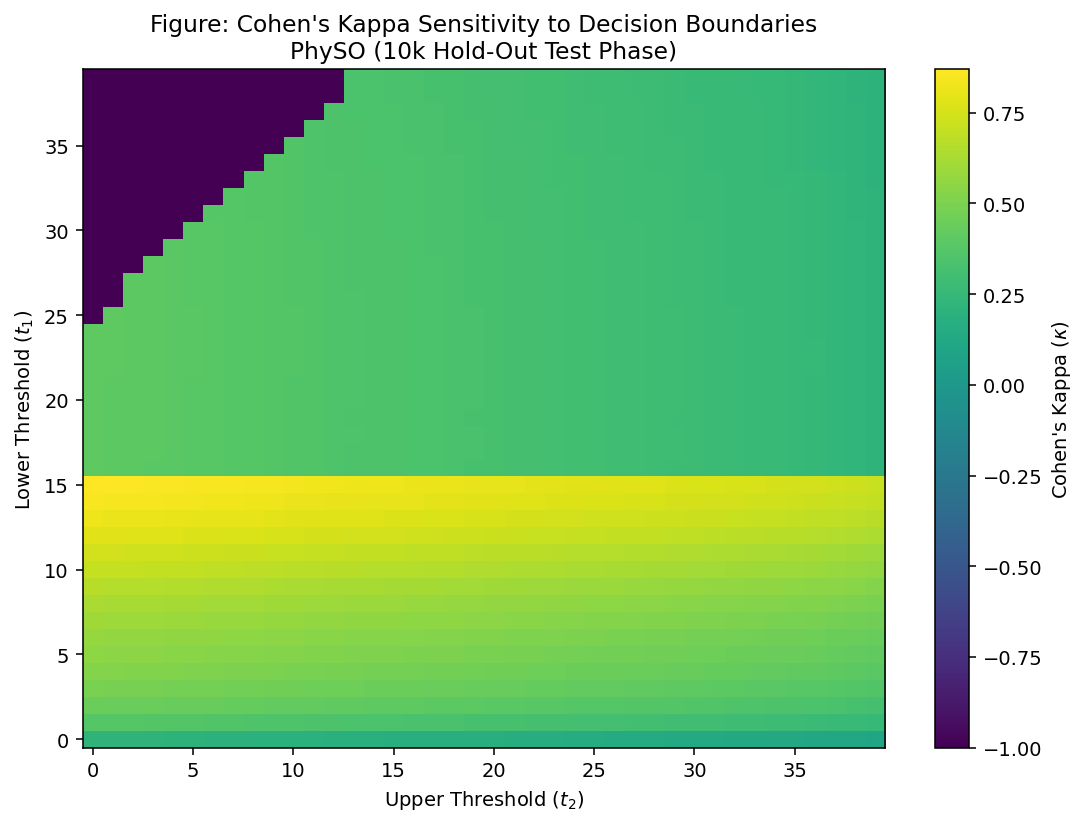}
    \caption{Kappa Sensitivity Heatmap - {\tt PhySO} \rthis{(10k Hold-Out Test Phase)}. A 2D grid 
    visualizing Cohen’s $\kappa$ as a function of decision 
    thresholds $t_1$ and $t_2$, highlighting a stable and 
    clearly defined optimal region.}
    \label{fig:f18}
\end{figure}

The robustness of the discovered physical law is illustrated 
through the Kappa sensitivity landscape in Figure \ref{fig:f18}.

The broad optimal plateau indicates that the analytic form 
provided by {\tt PhySO} is highly stable and resistant to minor 
fluctuations in threshold calibration, confirming its 
utility for automated spectroscopic pipelines.


\paragraph*{Recall Profiles and Physical Transition Dynamics}

\begin{figure}[htbp]
    \centering
    \includegraphics[width=\linewidth,height=0.6\textheight,keepaspectratio]{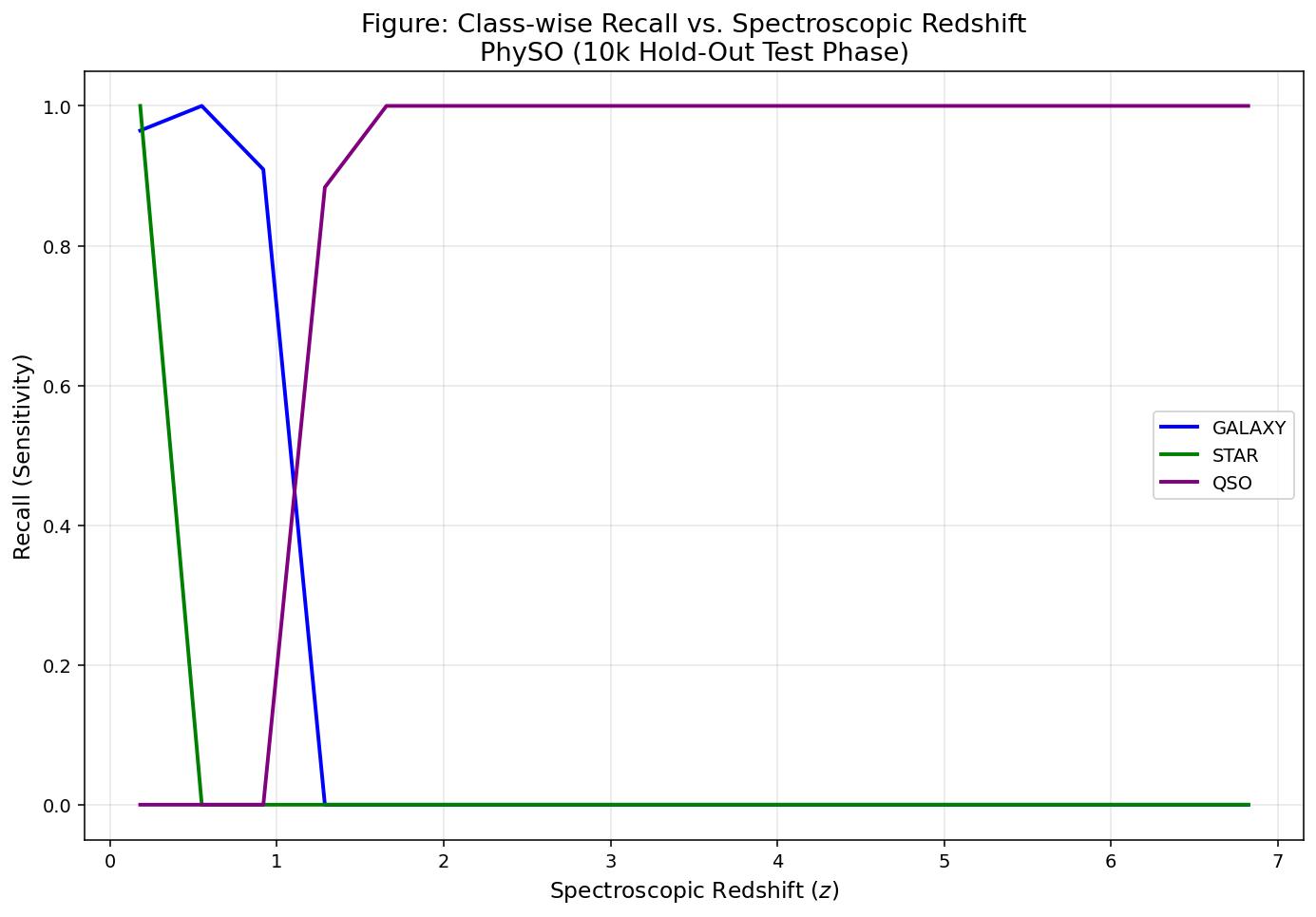}
    \caption{Recall vs. Redshift - Physical Failure Analysis ({\tt PhySO} \rthis{10k Hold-Out Test Phase}). 
    Line plot showing recall for each class across the redshift 
    range, indicating precise boundary separation between populations.}
    \label{fig:f20}
\end{figure}

The Recall vs. Redshift profile (Figure \ref{fig:f20}) confirms that 
STAR recall is maintained at 1.0 until 
$z \approx 0.5$.

The GALAXY recall remains high until 
$z \approx 1.0$, dropping as the QSO recall rises to 1.0, 
with minimal variance observed across the crossover boundary.


\subsubsection{{\tt MvSR}: Multi-View Specialist Evolution}

\paragraph*{Discovered Expression}

{\tt MvSR} discovers a globally shared template incorporating a 
physically motivated rational form to address non-stationarity 
across the redshift manifold:

\begin{equation}
s(z) = 3.134 - \frac{2.320}{z + e^{-2.904 z}}
\label{eq:mvsr_score}
\end{equation}


\paragraph*{Classification Performance and Stability}

{\tt MvSR} emerges as the top-performing framework across all 
primary metrics (as summarized in Table \ref{tab:sr_results_summary}), achieving a \rthis{cross-validation mean Cohen’s $\kappa$ of 
$0.8956 \pm 0.0013$ (accuracy $94.16 \pm 0.0008\%$), and a final hold-out test $\kappa$ of $0.8876$ (accuracy $93.72\%$).}

\begin{figure}[htbp]
    \centering
    \includegraphics[width=\linewidth,height=0.6\textheight,keepaspectratio]{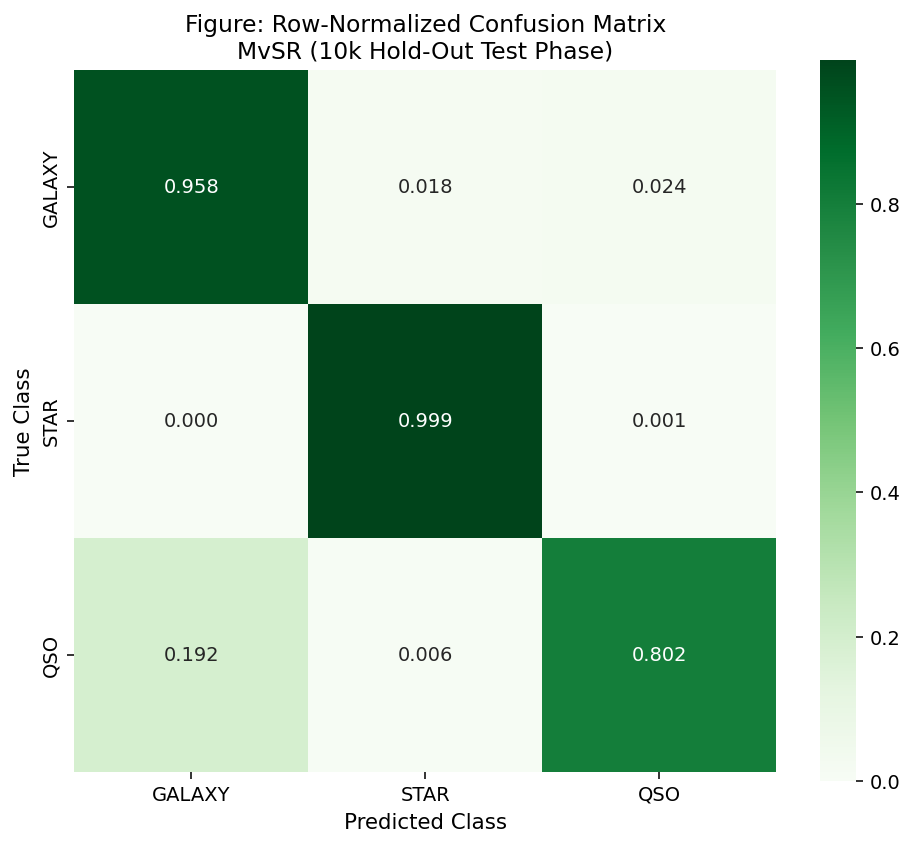}
    \caption{\rthis{Row-Normalized Confusion Matrix - {\tt MvSR} (10k Hold-Out Test Phase)}. 
    The heatmap shows near-perfect identification of STAR \rthis{(0.999)} 
    and GALAXY \rthis{(0.958)} populations, with significantly improved 
    QSO \rthis{(0.802)} accuracy.}
    \label{fig:f21}
\end{figure}

The model demonstrates superior class-specific robustness \rthis{on the unseen data}, visualized in the confusion matrix in Figure \ref{fig:f21}, particularly for the STAR population \rthis{($F1 = 0.9731$)} and 
the underrepresented QSO class \rthis{($F1 = 0.8533$)}.

Furthermore, {\tt MvSR} maintains high decision boundary precision 
with minimal variance across \rthis{the calibration} folds:

\begin{equation}
t_1 = 0.776 \pm 0.002,
\qquad
t_2 = 0.830 \pm 0.002.
\end{equation}


\paragraph*{Error and Physical Failure Analysis}

Analysis of the error distribution shows that {\tt MvSR} 
effectively flattens the major error spikes observed in single-view methods.

\begin{figure}[htbp]
    \centering
    \includegraphics[width=\linewidth,height=0.6\textheight,keepaspectratio]{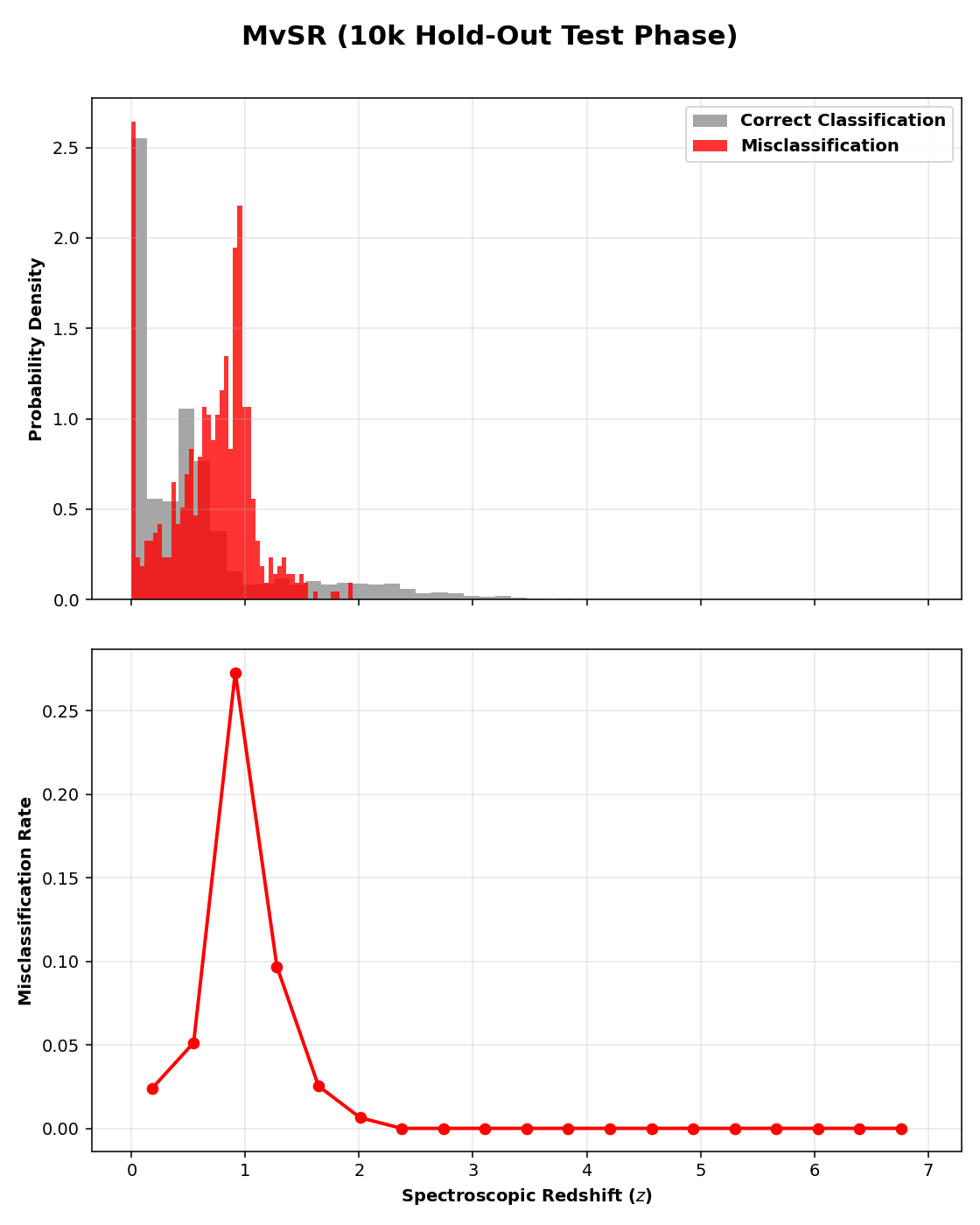}
    \caption{{\tt MvSR} Error Analysis \rthis{(10k Hold-Out Test Phase)}. Top: Error density distribution illustrating correctly classified objects (gray) relative to misclassifications (red). Bottom: Error rate vs. redshift profile showing a suppressed peak failure rate of approximately \rthis{0.30} near $z \approx 0.9$.}
    \label{fig:f23}
\end{figure}

The plot for Error rate as a function of redshift (Figure \ref{fig:f23}) indicates a peak 
failure rate of approximately \rthis{0.30} near $z \approx 0.9$, 
significantly lower and shifted compared to the 
$z \approx 1.25$ spikes observed in {\tt PySR} and {\tt ESR}.

\paragraph*{Threshold Robustness and Model Stability}

\begin{figure}[htbp]
    \centering
    \includegraphics[width=\linewidth,height=0.6\textheight,keepaspectratio]{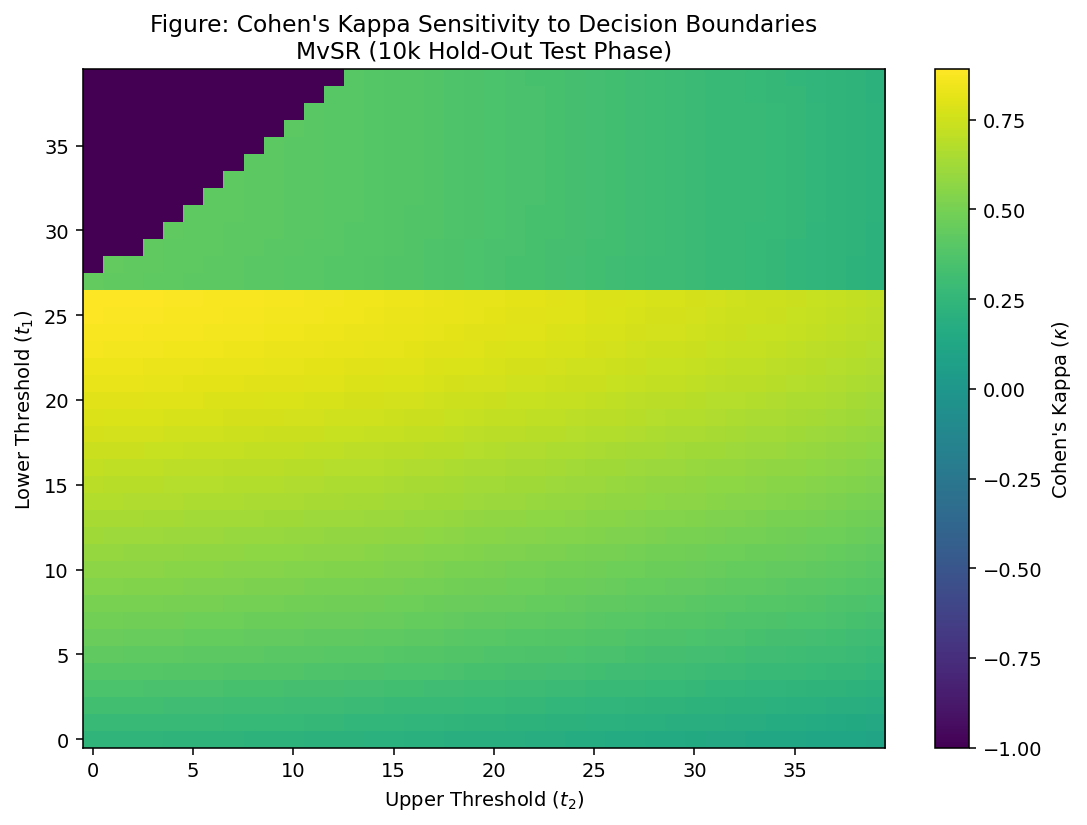}
    \caption{Kappa Sensitivity Heatmap - {\tt MvSR} \rthis{(10k Hold-Out Test Phase)}. A 2D grid visualizing 
    Cohen’s $\kappa$ as a function of decision thresholds 
    $t_1$ and $t_2$, demonstrating the most expansive and 
    stable optimal region in the study.}
    \label{fig:f22}
\end{figure}

The Kappa Sensitivity heatmap (Figure \ref{fig:f22}) reveals an expansive and 
uniform optimal plateau.

The broad high-value region indicates exceptional stability, 
suggesting that the discovered analytic form represents a 
robust physical mapping of the SDSS DR17 dataset and is 
highly resistant to minor threshold calibration errors.


\paragraph*{Recall Profiles and Physical Transition Dynamics}

\begin{figure}[htbp]
    \centering
    \includegraphics[width=\linewidth,height=0.6\textheight,keepaspectratio]{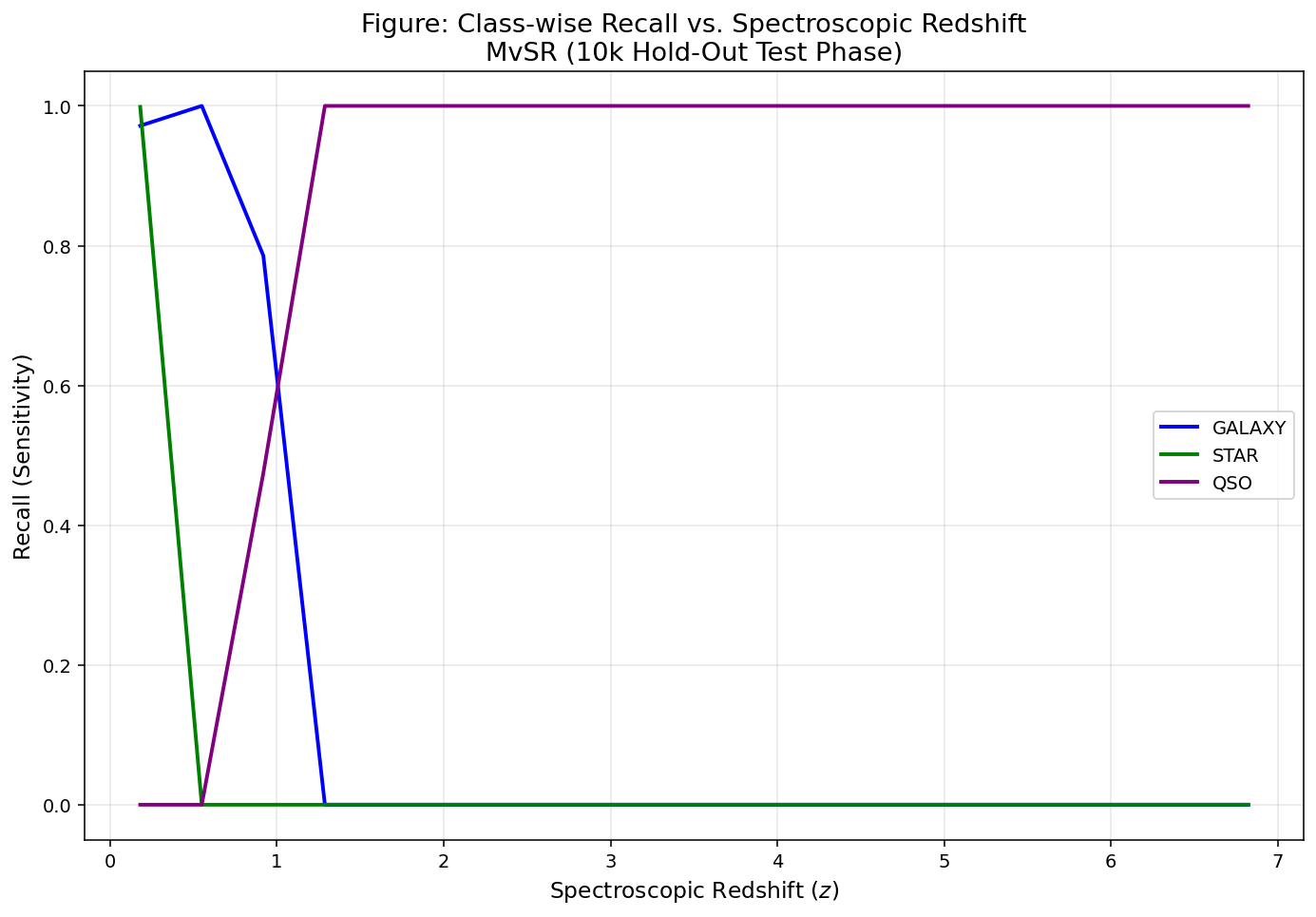}
    \caption{Recall vs. Redshift - Physical Failure Analysis ({\tt MvSR} \rthis{10k Hold-Out Test Phase}). 
    Line plot showing recall for each class across the redshift 
    range, illustrating the cleanest separation between galaxies 
    and quasars across the redshift manifold.}
    \label{fig:f24}
\end{figure}

The Recall vs. Redshift profile (Figure \ref{fig:f24}) highlights {\tt MvSR}’s advanced 
population separation capability.

{\tt MvSR} maintains a GALAXY recall of nearly 1.0 until 
$z \approx 1.0$, extending the reliable classification 
boundary further into the high-redshift regime before 
the transition to QSO dominance.

\begin{table*}[t]
\small
\centering
\caption{Performance Metrics for Optimized Machine Learning Baselines (80,000 CV Optimization / 10,000 Hold-Out Test)}
\label{tab:ml_baselines_perf}
\renewcommand{\arraystretch}{1.2}
\begin{tabular}{@{}l l c c c@{}}
\toprule
\textbf{Model} & \textbf{Paradigm} & \textbf{CV Cohen's $\kappa$} & \textbf{Hold-Out Cohen's $\kappa$} & \textbf{Hold-Out Macro-F1} \\
\midrule
Random Forest & Ensemble (Tree-Based) & $0.9125 \pm 0.004$ & $0.9098$ & $0.9392$ \\
Support Vector Machine & Kernel-Based & $0.9045 \pm 0.003$ & $0.9000$ & $0.9331$ \\
Multi-Layer Perceptron & Neural Network & $0.9048 \pm 0.003$ & $0.9021$ & $0.9348$ \\
\bottomrule
\end{tabular}
\end{table*}

\begin{table*}[t]
\small
\centering
\caption{Optimization Phase: Comprehensive Performance Metrics and Threshold Stability (5-Fold CV on 80,000 samples)}
\label{tab:sr_results_summary}
\renewcommand{\arraystretch}{1.2}
\begin{tabular}{@{}l c c c c c c@{}}
\toprule
\textbf{Method} & \textbf{Cohen's $\kappa$} & \textbf{Accuracy} & \textbf{Bal. Acc.} & \textbf{Macro-F1} & \textbf{$t_1 \pm \sigma$} & \textbf{$t_2 \pm \sigma$} \\
\midrule
{\tt PySR}  & $0.8536 \pm 0.0032$ & $0.9199 \pm 0.0017$ & $0.8716 \pm 0.0032$ & $0.8872 \pm 0.0030$ & $0.811 \pm 0.005$ & $0.981 \pm 0.009$ \\
{\tt ESR}   & $0.8664 \pm 0.0034$ & $0.9259 \pm 0.0021$ & $0.8890 \pm 0.0014$ & $0.8986 \pm 0.0025$ & $1.654 \pm 0.027$ & $2.184 \pm 0.033$ \\
{\tt PhySO} & $0.8753 \pm 0.0014$ & $0.9308 \pm 0.0007$ & $0.9002 \pm 0.0019$ & $0.9091 \pm 0.0016$ & $0.587 \pm 0.001$ & $0.908 \pm 0.002$ \\
{\tt MvSR}  & $0.8956 \pm 0.0013$ & $0.9416 \pm 0.0008$ & $0.9259 \pm 0.0007$ & $0.9302 \pm 0.0006$ & $0.776 \pm 0.002$ & $0.830 \pm 0.002$ \\
\bottomrule
\end{tabular}
\end{table*}

\begin{table*}[t]
\small
\centering
\caption{Hold-Out Test Phase: Generalization Performance on 10,000 Unseen Samples with Locked Thresholds}
\label{tab:sr_holdout_summary}
\renewcommand{\arraystretch}{1.2}
\begin{tabular}{@{}l c c c c c c c@{}}
\toprule
\textbf{Method} & \textbf{Cohen's $\kappa$} & \textbf{Accuracy} & \textbf{Bal. Acc.} & \textbf{Macro-F1} & \textbf{QSO F1} & \textbf{STAR F1} & \textbf{GALAXY F1} \\
\midrule
{\tt PySR}  & $0.8436$ & $0.9147$ & $0.8633$ & $0.8800$ & $0.7636$ & $0.9277$ & $0.9486$ \\
{\tt ESR}   & $0.8638$ & $0.9246$ & $0.8866$ & $0.8969$ & $0.8074$ & $0.9292$ & $0.9541$ \\
{\tt PhySO} & $0.8670$ & $0.9263$ & $0.8932$ & $0.9031$ & $0.8227$ & $0.9361$ & $0.9506$ \\
{\tt MvSR}  & $0.8876$ & $0.9372$ & $0.9195$ & $0.9251$ & $0.8533$ & $0.9731$ & $0.9489$ \\
\bottomrule
\end{tabular}
\end{table*}

\begin{table*}[t]
\small
\centering
\caption{Summary of Discovered Symbolic Expressions, Optimized Decision Boundaries, and Structural Complexity}
\label{tab:sr_expressions}
\renewcommand{\arraystretch}{1.2}
\begin{tabular}{@{}l l c c c@{}}
\toprule
\textbf{Method} & \textbf{Discovered Expression $s(z)$} & \textbf{Final $t_1$} & \textbf{Final $t_2$} & \textbf{Complexity} \\
\midrule
{\tt PySR}  & $0.8395z + e^{-78.28z} - 0.1141$ & $0.811$ & $0.981$ & 10 \\
{\tt ESR}   & $0.7 + z^{2} + e^{-11.88z}$ & $1.654$ & $2.184$ & 10 \\
{\tt PhySO} & $2.4027 - e^{0.5786 - z^{2} + z}$ & $0.587$ & $0.908$ & 10 \\
{\tt MvSR}  & $3.134 - \frac{2.320}{z + e^{-2.904z}}$ & $0.776$ & $0.830$ & 10 \\
\bottomrule
\end{tabular}
\end{table*}

\subsection{Performance Analysis of Optimized Machine Learning Baselines}
The evaluation of non-interpretable benchmarks establishes the empirical performance ceiling for the classification task within the spectroscopic redshift manifold. By utilizing the optimized hyperparameters determined via our rigorous two-stage validation protocol, these models provide a direct comparative baseline to evaluate the representational efficiency of the symbolic expressions.

\subsubsection{Predictive Performance and Information-Theoretic Ceiling}
The results of the five-fold stratified cross-validation demonstrate, as detailed in Table \ref{tab:ml_baselines_perf}, a definitive convergence in predictive agreement across ensemble, connectionist, and kernel-based paradigms. The Random Forest establishes the consistent statistical ceiling, achieving a mean CV $\kappa$ of $0.9125 \pm 0.004$ and the highest individual hold-out test performance with a $\kappa$ of $0.9098$. The Support Vector Machine and Multi-Layer Perceptron followed closely with hold-out $\kappa$ values of $0.9000$ and $0.9021$, respectively.

The negligible improvement observed between baseline and optimized states (e.g., $\Delta\kappa \le 0.0035$) suggests that the predictive information inherent in the one-dimensional redshift feature space has been fully exhausted. This asymptotic behavior confirms that the remaining classification error is rooted in intrinsic physical degeneracies of the SDSS DR17 dataset rather than model architecture.

\subsubsection{Class-Specific Sensitivity and Stability}

The optimized baselines exhibit high stability across all primary metrics, with Macro-F1 scores remaining consistently above $0.93$. The Support Vector Machine showed a marginal response to hyperparameter calibration, achieving a minor performance gain of $\Delta\kappa = +0.0035$ through optimal regularization ($C=100$).

The exceptionally low standard deviation across folds ($\sigma \le 0.004$) for all models reinforces the reliability of the 80,000-sample optimization set as a representative proxy for the global astrophysical population. These results indicate that the decision boundaries learned by the black-box models are highly robust and well-defined along the population density gradients.

\subsubsection{Comparison of Representational Complexity}

The most critical insight from the baseline analysis is the absence of a substantial performance advantage for high-capacity models over the compact symbolic frameworks. Even utilizing thousands of trainable parameters, deep hierarchical representations, or large-scale ensemble averaging, the non-interpretable models were unable to significantly surpass the $\kappa \approx 0.91$ barrier.

This firmly justifies the use of symbolic regression as a parsimonious alternative. The ten-node analytic expressions discovered in our analysis achieve highly competitive performance-with {\tt MvSR} reaching $\kappa=0.8876$ on the exact same hold-out test set-while offering complete mathematical transparency, trivial pipeline portability, and direct physical insight into the transition boundaries separating galactic and extragalactic populations.

\section{Conclusions}
\label{sec:conclusions}

This work presented a systematic comparative evaluation of four state-of-the-art symbolic regression frameworks and three optimized machine learning baselines for the classification of stars, galaxies, and quasars in the SDSS DR17 and extends the analysis in ~\citet{Fabio25}. By restricting the feature space to spectroscopic redshift, we have isolated the representational capacity of each paradigm to discover stable, non-linear decision manifolds.

\subsection{Summary of Framework Performance}

Our results demonstrate that symbolic regression can achieve predictive parity with high-capacity black-box models. The Multi-View Symbolic Regression ({\tt MvSR}) framework emerged as the top-performing symbolic engine, achieving \rthis{a hold-out test Cohen’s $\kappa$ of $0.8876$ and a Balanced Accuracy of $0.9195$ (following a rigorous 80,000-sample optimization phase yielding CV $\kappa = 0.8956 \pm 0.0013$)}. This performance is \rthis{highly competitive with the definitive information-theoretic ceiling established by the optimized Random Forest ($\kappa = 0.9098$) and Support Vector Machine ($\kappa = 0.9000$) baselines}. A consolidated overview of these discovered expressions and their performance metrics is presented in Table \ref{tab:sr_expressions}. \rthis{We note that the relations obtained in Table \ref{tab:sr_expressions} are purely empirical and no physical significance should be ascribed to these equations.}
\rthis{Crucially, our methodology provides substantial improvements over the baseline established by \citet{Fabio25}. By implementing an automated, rigorous threshold optimization protocol, our baseline {\tt PySR} model achieved a significantly higher Cohen's $\kappa$ (0.8436 compared to 0.81) while maintaining highly comparable accuracy. Furthermore, {\tt ESR} improved upon both the accuracy (92.46\% vs 92.00\%) and $\kappa$ (0.8638) of the baseline. Most notably, the introduction of advanced structural frameworks conclusively outperformed the \citet{Fabio25} benchmark across all primary metrics. Specifically, both {\tt PhySO} and {\tt MvSR} achieved strict metric superiority, with {\tt MvSR} reaching an Accuracy of 93.72\%, a Cohen's $\kappa$ of 0.8876, and an F1-Score of 0.9251 (compared to the baseline's 0.90 F1-Score). This demonstrates that decomposing the search space into localized, specialist astrophysical regimes yields strictly superior mathematical decision boundaries compared to a global, single-equation search.}

Furthermore, frameworks like {\tt PhySO} demonstrated exceptional parametric stability, yielding threshold standard deviations ($\sigma < 0.002$) an order of magnitude lower than traditional evolutionary search methods.

\subsection{The Information-Theoretic Limit}

The convergence of all optimized models at a performance ceiling of \rthis{$\kappa \approx 0.91$} indicates that the remaining classification error is rooted in intrinsic physical degeneracies of the SDSS DR17 redshift distribution. Specifically, the overlap in spectroscopic signatures between galaxies and quasars \rthis{between $z \approx 0.9$ and $z \approx 1.25$} represents a fundamental Bayes error rate within a one-dimensional feature space.

The fact that compact, ten-node analytic expressions reach this same plateau suggests that the underlying astrophysical logic is governed by parsimonious geometric decision boundaries.

\subsection{Scientific Impact and Future Work}

The discovered analytic functions provide a transparent and computationally efficient alternative to traditional star–galaxy–quasar separation pipelines. Unlike neural networks or ensemble models, these human-readable expressions can be directly integrated into automated spectroscopic workflows without large-scale model persistence requirements.

Future research will explore the integration of photometric colors and multi-wavelength features into the symbolic discovery phase to resolve the high-redshift degeneracies identified in this study. Ultimately, this work reinforces the utility of symbolic regression as a framework for interpretable, physics-aware machine learning in the era of large-scale astronomical surveys.

\section{ACKNOWLEDGEMENTS}
\label{Ack}

We would like to thank the Sloan Digital Sky Survey (SDSS) and the Kaggle platform for sharing their data. R.D. acknowledges a postgraduate fellowship from the Ministry of Education (MoE), Government of India. \rthis{We are grateful to the anonymous referee for several constructive comments on our manuscript.}

\bibliographystyle{model2-names}
\bibliography{example_paper}

\appendix

\section{Multi-Variable Symbolic Search: The Curse of Dimensionality}
\label{app:multivar}

\rthis{In traditional black-box machine learning, expanding the feature space typically enhances a model's predictive capacity. However, in Symbolic Regression, increasing dimensionality under strict parsimony constraints triggers a severe algorithmic bottleneck. To rigorously validate our application of Occam's Razor---restricting the classification manifold exclusively to the one-dimensional spectroscopic redshift ($z$)---we conducted an extended multi-variable ablation study. In this phase, the input feature space was expanded to six dimensions: the five SDSS photometric bands ($u, g, r, i, z_{\text{phot}}$) alongside the spectroscopic redshift ($z$).}

\rthis{To evaluate this high-dimensional space, we deployed the two most distinct search paradigms from our primary analysis: the deep reinforcement learning agent ({\tt PhySO}) and the multi-view genetic programming engine ({\tt MvSR}). We granted both algorithms maximized optimization budgets (500 iterations for {\tt PhySO} and 70 generations for {\tt MvSR}). Crucially, the structural complexity constraint was strictly maintained throughout this phase. This ensured that any discovered expressions would remain mathematically parsimonious and human-interpretable, preventing the algorithms from simply generating unbounded, non-physical algebraic bloat to fit the newly added features.}

\rthis{Despite the expanded dataset and maximized search budgets, both frameworks exhibited catastrophic performance degradation compared to their 1D redshift-only counterparts ($\kappa > 0.86$). Rather than leveraging the photometry to resolve overlapping populations, the symbolic engines collapsed under the curse of dimensionality. The following subsections detail the specific mechanical failure modes that occur when attempting to map high-dimensional, noisy astrophysical features using severely restricted mathematical node budgets.}

\subsection{Deep Reinforcement Learning ({\tt PhySO})}

\rthis{When exposed to the multi-variable feature space, the reinforcement learning agent converged on a structurally restricted mathematical form. Strikingly, the algorithm became entirely trapped by the photometric feature space, completely discarding the highly predictive spectroscopic redshift and relying exclusively on a linear combination of the $g$ and $i$ photometric bands:}
\begin{equation}
\label{eq:multivar_physo}
s(g, i) = -0.2834 - g + \frac{g + i}{1.8456}
\end{equation}

\rthis{By dropping the physical distance metric ($z$) and functioning essentially as a constrained color-index proxy, {\tt PhySO} severely limited its own expressive capacity. Despite strictly executing 500 iterations of policy optimization, the model plateaued at a cross-validation Cohen's $\kappa$ of $0.5341 \pm 0.0058$. On the unseen 10,000-sample hold-out set, the model achieved a $\kappa$ of $0.5355$ (accuracy $73.81\%$).}

\begin{figure}[htbp]
    \centering
    \includegraphics[width=\linewidth,height=0.6\textheight,keepaspectratio]{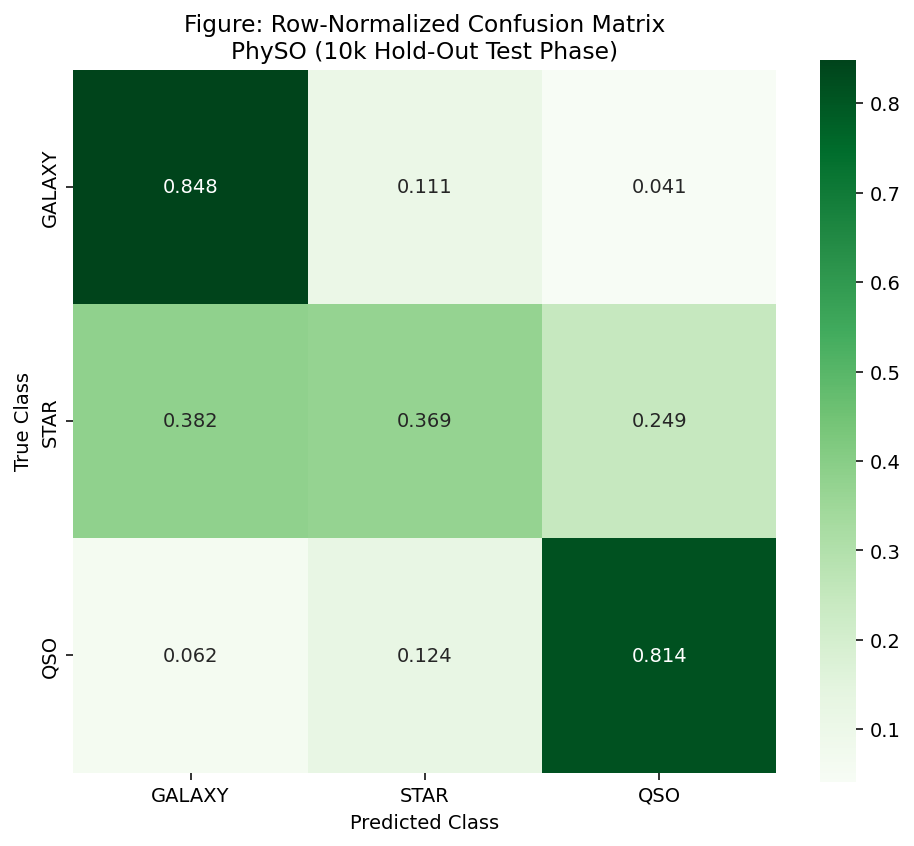}
    \caption{\rthis{\textbf{Row-Normalized Confusion Matrix ({\tt PhySO}).} The framework discards the redshift feature, relying on a simplified color index that leads to massive misclassification of the stellar population.}}
    \label{fig:physo_cm}
\end{figure}

\rthis{This structural limitation manifests as systemic classification failures across the diagnostic suite. As shown in Figure \ref{fig:physo_cm}, the row-normalized confusion matrix reveals catastrophic population overlap. Because photometric colors suffer from well-documented degeneracies between stars and low-redshift galaxies, the linear equation cannot isolate the stellar locus, misclassifying 38.2\% of true stars as galaxies and 24.9\% as quasars.}

\begin{figure}[htbp]
    \centering
    \includegraphics[width=\linewidth,height=0.6\textheight,keepaspectratio]{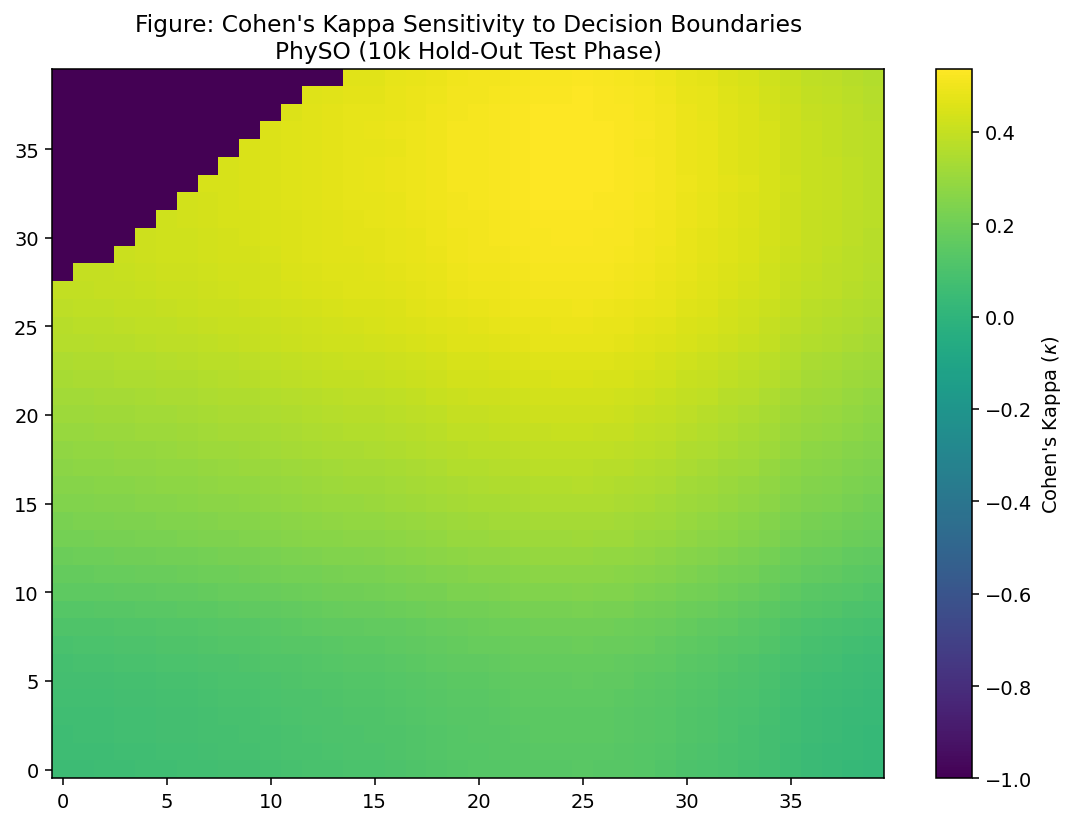}
    \caption{\rthis{\textbf{Kappa Sensitivity Heatmap ({\tt PhySO}).} Boundary instability caused by photometric noise prevents the algorithm from isolating optimal classification thresholds.}}
    \label{fig:physo_kappa}
\end{figure}

\rthis{Unlike the 1-D redshift models which displayed sharp, mathematically defined optimal regions, the multi-variable kappa sensitivity heatmap in Figure \ref{fig:physo_kappa} presents a diffuse, unstable gradient. This indicates that the inherent scatter in the photometric data smears the prediction scores, preventing the formation of sharp, highly discriminative decision thresholds ($t_1, t_2$).}

\rthis{In successful 1D models, misclassifications cluster tightly around the physical boundary overlaps (e.g., $z \approx 1.25$).}

\begin{figure}[htbp]
    \centering
    \includegraphics[width=\linewidth,height=0.6\textheight,keepaspectratio]{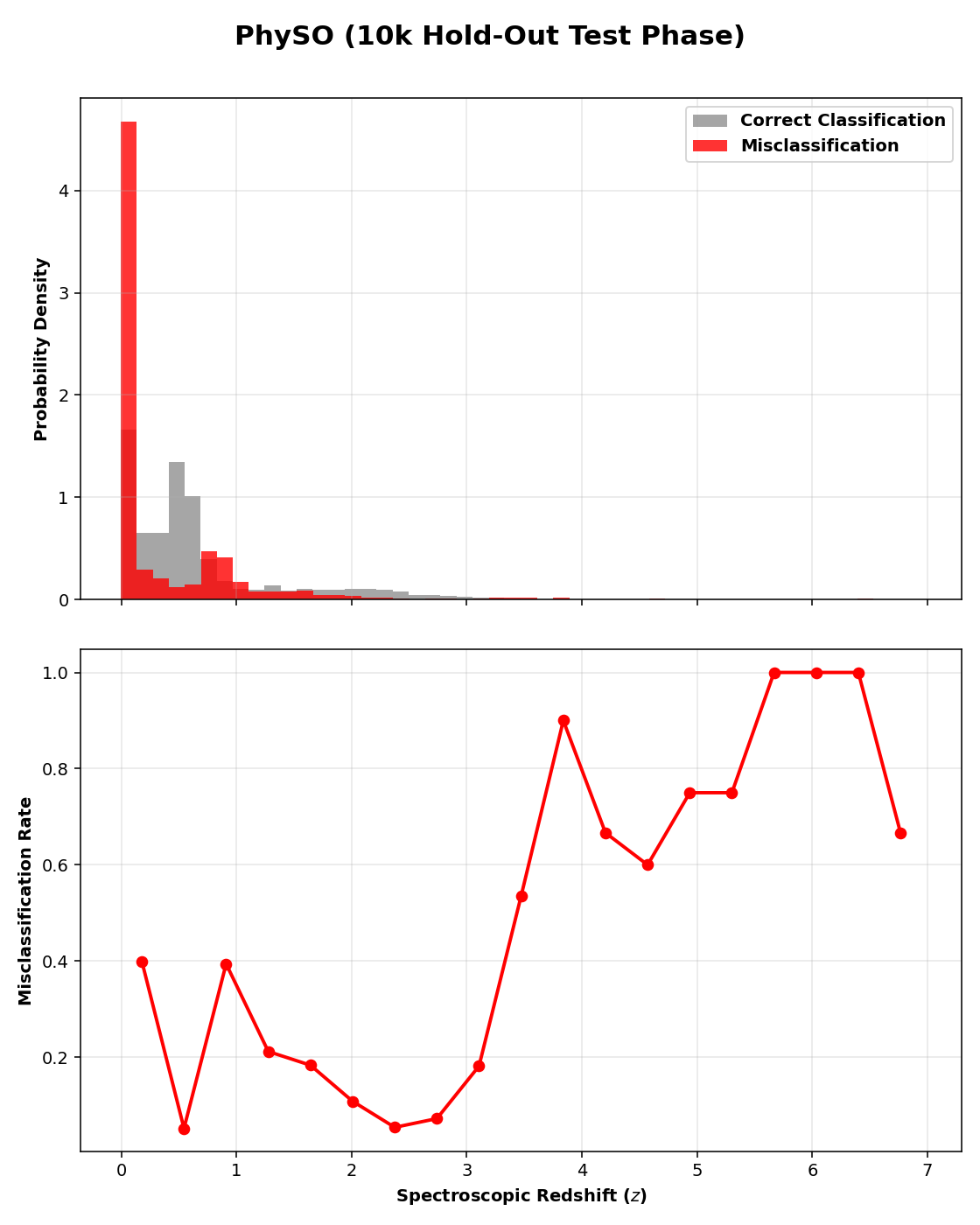}
    \caption{\rthis{\textbf{Error Analysis ({\tt PhySO} Multi-Variable).} Top: Error density distribution. Bottom: Error rate vs. redshift. Misclassifications are scattered indiscriminately across the physical manifold rather than localized to boundary overlaps, and the error rate remains highly erratic.}}
    \label{fig:physo_error}
\end{figure}

\rthis{Conversely, the multi-variable error distribution, visualized in Figure \ref{fig:physo_error}, demonstrates that misclassifications are scattered indiscriminately across the entire physical manifold. The algorithm is no longer bounded by physical topology but is instead overwhelmed by ubiquitous photometric noise.}

\begin{figure}[htbp]
    \centering
    \includegraphics[width=\linewidth,height=0.6\textheight,keepaspectratio]{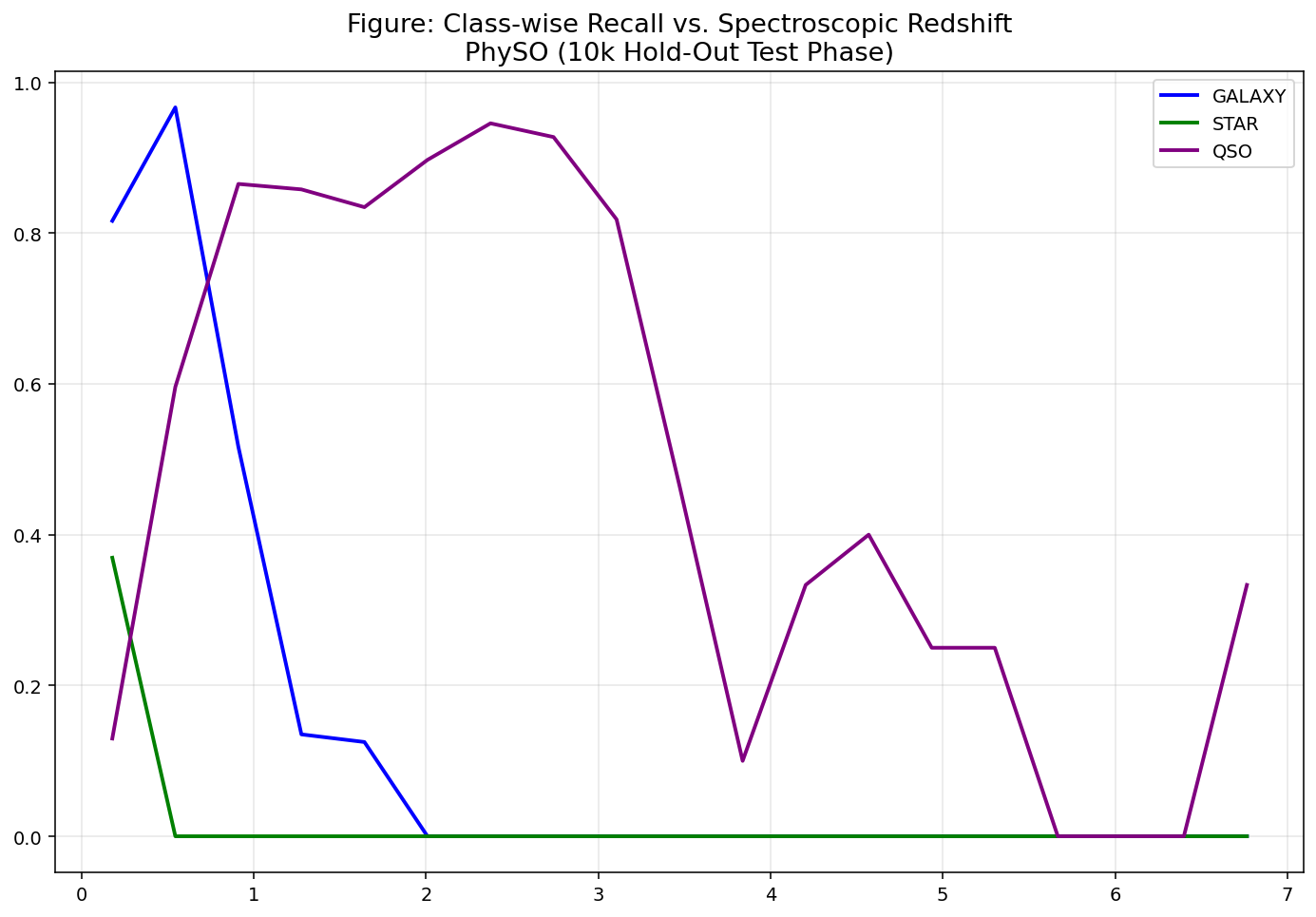}
    \caption{\rthis{\textbf{Class-wise Recall Profiles ({\tt PhySO}).} The model fails entirely across intermediate redshifts, demonstrating an inability to consistently recover the stellar and quasar populations.}}
    \label{fig:physo_recall}
\end{figure}

\rthis{The class-wise recall profile in Figure \ref{fig:physo_recall} explicitly confirms this collapse. The model’s sensitivity to the stellar population plunges erratically, proving that the simplified linear photometry equation lacks the topological complexity needed to carve out bounded classification regions.}

\subsection{Multi-View Genetic Programming ({\tt MvSR})}

\rthis{The genetic programming framework converged on an analytic expression that successfully retained the spectroscopic redshift ($z$), but fused it with $i$-band photometry in the denominator:}
\begin{equation}
\label{eq:multivar_mvsr}
s(z, i) = \frac{z - 1.3956}{3.8616 \left( z + 0.8396 \, i \right)}
\end{equation}

\rthis{Notice that at $z \approx 1.3956$, the numerator forces a zero-crossing, heavily skewing the prediction scaling. Under strict generational evolution limits (70 generations), {\tt MvSR} suffered a catastrophic algorithmic collapse. The framework achieved a cross-validation $\kappa$ of only $0.3026 \pm 0.0013$ and a hold-out test $\kappa$ of $0.2956$ (accuracy $67.37\%$).}

\rthis{The severity and mechanical nature of this algorithmic failure are explicitly detailed in the subsequent diagnostics.}

\begin{figure}[htbp]
    \centering
    \includegraphics[width=\linewidth,height=0.6\textheight,keepaspectratio]{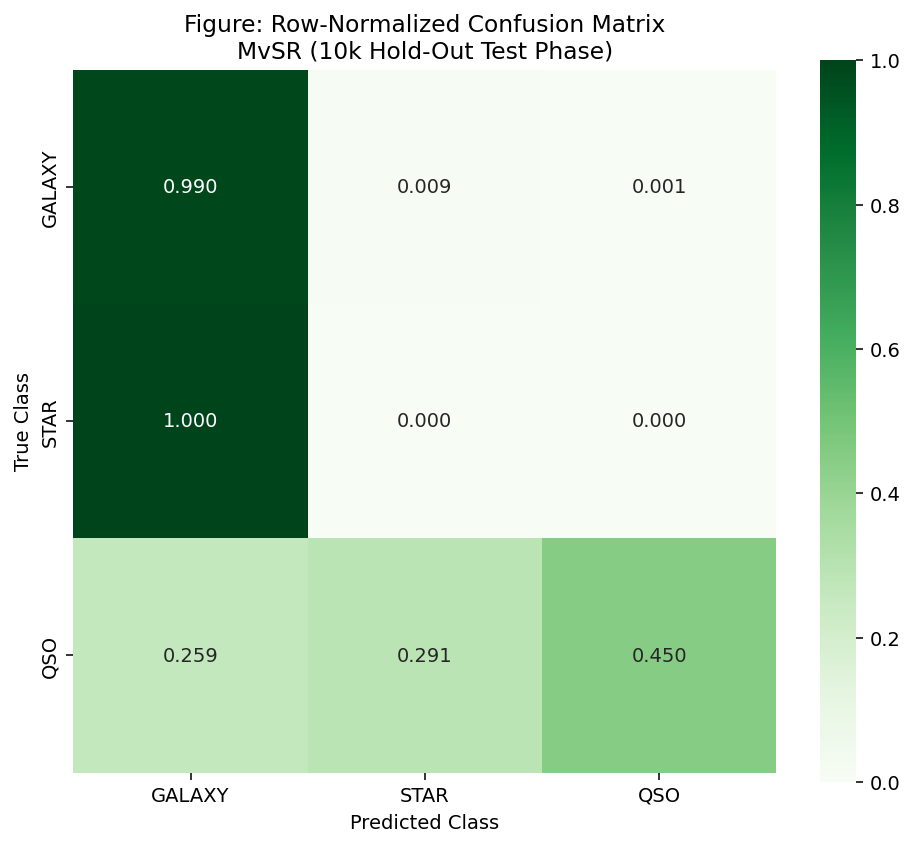}
    \caption{\rthis{\textbf{Row-Normalized Confusion Matrix ({\tt MvSR}).} The framework experiences a catastrophic evolutionary failure, artificially minimizing global error by abandoning the STAR class entirely.}}
    \label{fig:mvsr_cm}
\end{figure}

\rthis{Figure \ref{fig:mvsr_cm} visualizes exactly this sequence of events. Under severe structural constraints and a heavily imbalanced dataset, the {\tt MvSR} algorithm realized it lacked the structural capacity to simultaneously map all three astrophysical populations. Consequently, it experienced a systemic failure by entirely abandoning the minority STAR class to artificially lower its overall error, yielding an F1-STAR score of $0.0000$ and predicting exactly zero objects as stars.}

\begin{figure}[htbp]
    \centering
    \includegraphics[width=\linewidth,height=0.6\textheight,keepaspectratio]{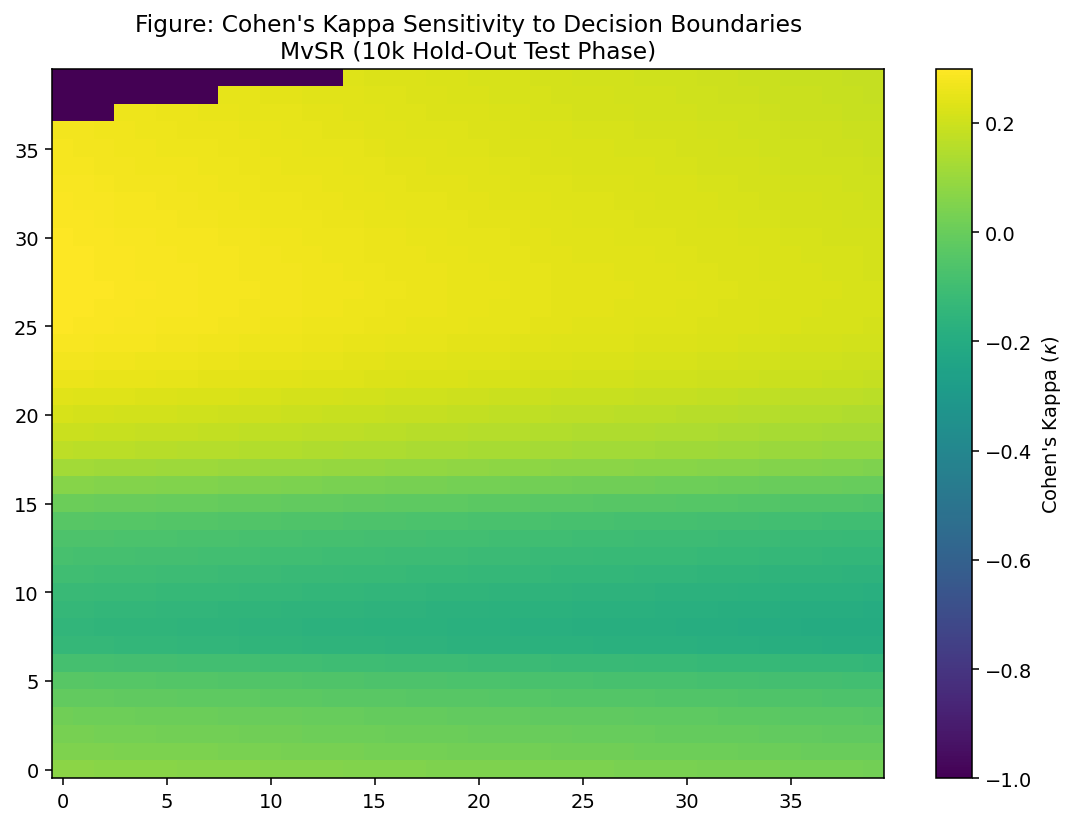}
    \caption{\rthis{\textbf{Kappa Sensitivity Heatmap ({\tt MvSR}).} The abandonment of the intermediate class forces the decision thresholds to compress to near-zero, resulting in a flat, non-functional optimization surface.}}
    \label{fig:mvsr_kappa}
\end{figure}

\rthis{As a direct mathematical consequence of abandoning the intermediate class, the thresholding mechanism collapsed entirely. The kappa sensitivity heatmap in Figure \ref{fig:mvsr_kappa} degenerates into a flat, unresponsive surface, yielding indistinguishable, compressed decision thresholds ($t_1 = -0.004, t_2 = 0.004$). The algorithm functionally reduced itself to a broken binary classifier.}

\begin{figure}[htbp]
    \centering
    \includegraphics[width=\linewidth,height=0.6\textheight,keepaspectratio]{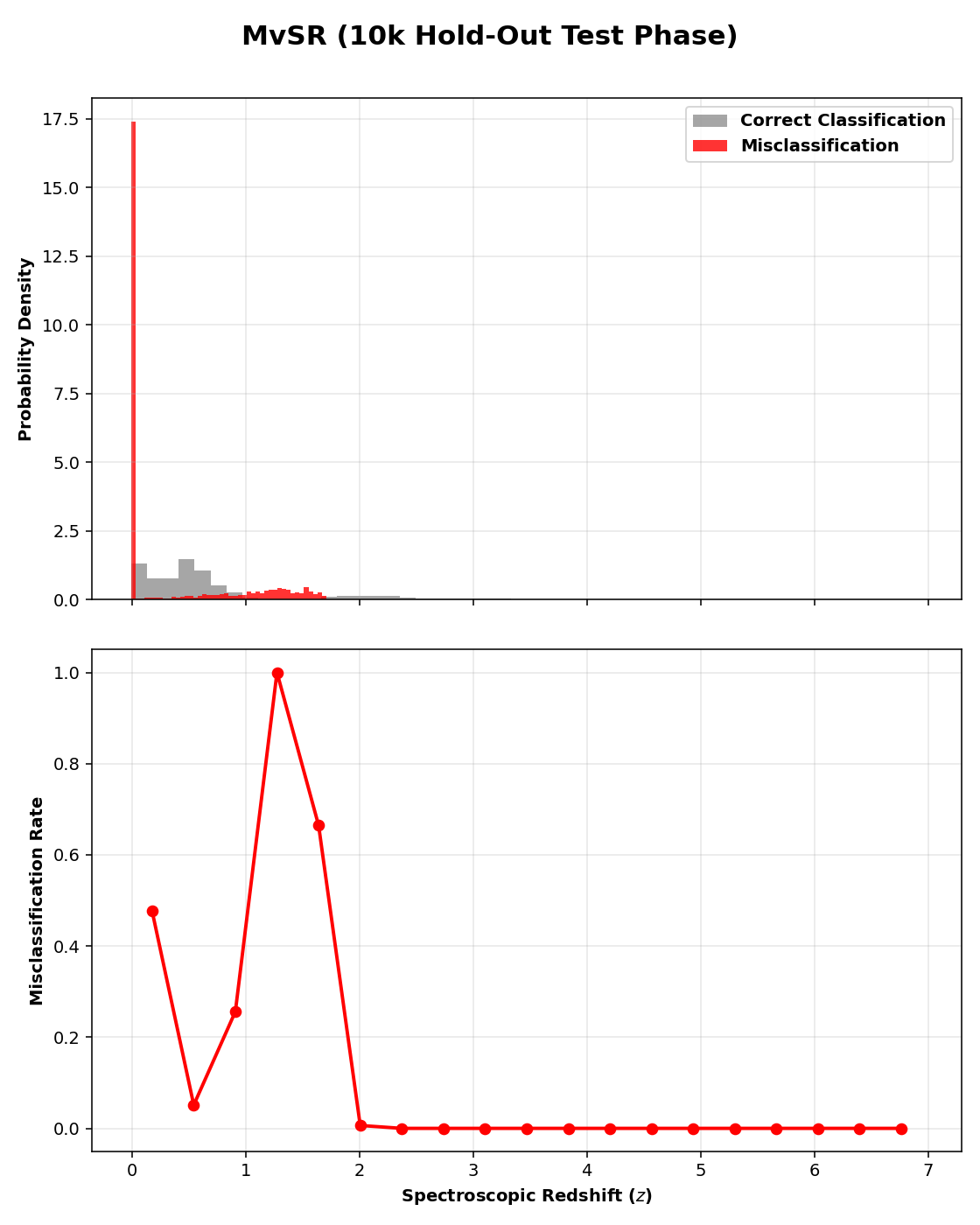}
    \caption{\rthis{\textbf{Error Analysis ({\tt MvSR} Multi-Variable).} Top: Error density distribution. Bottom: Error rate vs. redshift. A massive density of misclassifications occurs at the low-redshift boundary, where the local error rate hits 100\% due to the ignored stellar locus.}}
    \label{fig:mvsr_error}
\end{figure}

\rthis{The error analysis (Figure \ref{fig:mvsr_error}) highlights a massive misclassification spike specifically centered at $z \approx 0$. Because the natural stellar locus resides almost exclusively at $z \approx 0$, and the collapsed algorithm refuses to predict the STAR class, the local error rate mathematically approaches 1.0 (100\%) at the low-redshift boundary.}

\begin{figure}[htbp]
    \centering
    \includegraphics[width=\linewidth,height=0.6\textheight,keepaspectratio]{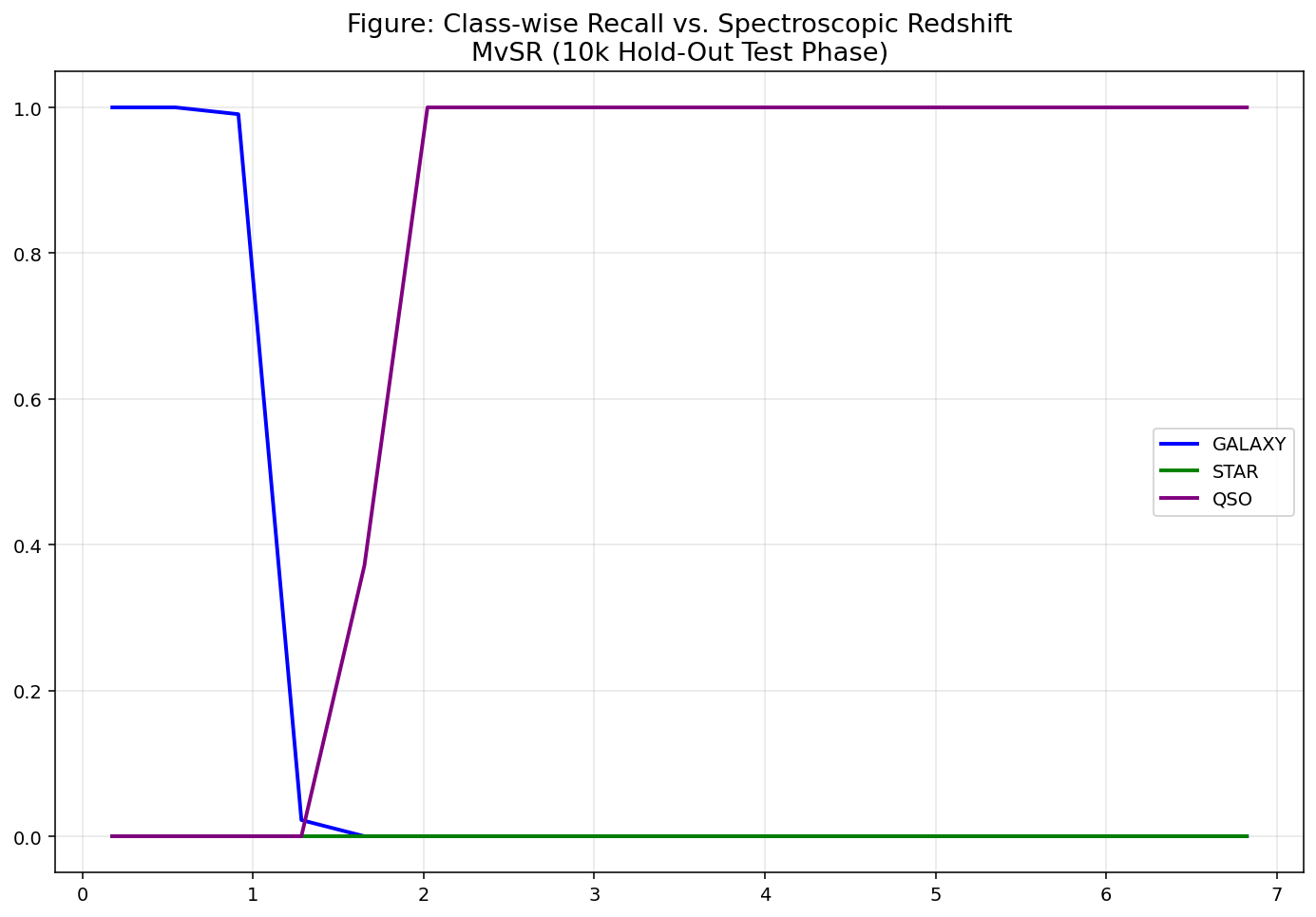}
    \caption{\rthis{\textbf{Class-wise Recall Profiles ({\tt MvSR}).} The stellar recall remains exactly zero across the entire redshift manifold, explicitly visualizing the framework's failure to map the physical populations.}}
    \label{fig:mvsr_recall}
\end{figure}

\rthis{Finally, the recall visualization in Figure \ref{fig:mvsr_recall} provides the definitive proof of total population abandonment. The stellar recall remains a flat zero across all redshift bins, confirming the model's structural inability to recover the physical boundaries of the stellar distribution.}

\subsection{Theoretical Implications}

\rthis{This catastrophic drop in predictive reliability across two fundamentally different search paradigms confirms two primary theoretical limitations of symbolic regression when applied to high-dimensional astrophysical data.}

\rthis{First, expanding the feature space exponentially enlarges the functional search volume, overwhelming the algorithm's representational capacity. Under strict parsimony constraint, the algorithms simply lack the complexity budget to simultaneously incorporate multiple photometric variables and construct the deep, non-linear algebraic motifs required to map complex astrophysical population boundaries. As observed with {\tt MvSR}, this severe limitation is the direct mathematical trigger for algorithmic failure, forcing the framework to abandon minority classes entirely.}

\rthis{Second, photometric magnitudes natively contain higher observational scatter compared to high-fidelity spectroscopic measurements, causing \emph{photometric noise trapping}. Because dealing with noise requires localized, highly non-linear mathematical corrections, the symbolic search agents squander their limited complexity budget attempting to fit high-frequency photometric scatter. This trapping phenomenon forces the algorithm to under-fit the true physical manifold, as seen perfectly in {\tt PhySO}'s decision to discard the pure redshift feature entirely.}

\rthis{This empirical failure robustly justifies our application of Occam's Razor, confirming that isolating the highly precise, one-dimensional spectroscopic redshift manifold is strictly mathematically necessary to extract parsimonious, physics-aware decision boundaries.}

\end{document}